\documentclass[useAMS,usenatbib]{mn2e}
\usepackage{epsfig}

\def\msun{{\rm {M}}_{\odot}}
\newcommand{\etal}{{et al.}~}
\newcommand{\eg}{{e.g.~}}
\newcommand{\ie}{{i.e.~}}

\def \ltsima{$\; \buildrel < \over \sim \;$}
\def \simlt{\lower.5ex\hbox{\ltsima}}            
\def \gtsima{$\; \buildrel > \over \sim \;$}
\def \gtsima{\mbox{$\; \buildrel > \over \sim \;$}}
\def \simgt{\lower.5ex\hbox{\gtsima}}            

\newcounter{cureqno}

\title[halo mass function]{
The halo mass function from the dark ages through the present day}
\author[Reed \etal] {
Darren S. Reed$^{1}$ \thanks{Email:
d.s.reed@durham.ac.uk},
Richard Bower$^{1}$, 
Carlos S. Frenk$^{1}$,
\newauthor
~~~~~~Adrian Jenkins$^{1}$, 
and Tom Theuns$^{1,2}$\\
$^1$Institute for Computational Cosmology, Dept. of Physics, 
University of Durham,  South Road, Durham DH1 3LE, UK\\ 
$^2$Dept. of Physics, Univ. of Antwerp, Campus Groenenborger, 
Groenenborgerlaan 171, B-2020 Antwerp, Belgium}

\pagerange{\pageref{firstpage}--\pageref{lastpage}}
\pubyear{2003}

\begin{document}

\maketitle

\label{firstpage}

\begin{abstract}


We use an array of high-resolution N-body simulations to determine the
mass function of dark matter haloes at redshifts 10-30.  We develop a
new method for compensating for the effects of finite simulation
volume that allows us to find an approximation to the true ``global''
mass function.  By simulating a wide range of volumes at different
mass resolution, we calculate the abundance of haloes of mass
$10^{5-12}$ $h^{-1} \msun$.  This enables us to predict accurately the
abundance of the haloes that host the sources that reionize the
universe.  In particular, we focus on the small mass haloes ($\simgt
10^{5.5-6} h^{-1} \msun$) likely to harbour population III stars where
gas cools by molecular hydrogen emission, early galaxies in which
baryons cool by atomic hydrogen emission at a virial temperature of
$\sim 10^{4}$K ($\sim 10^{7.5-8} h^{-1} \msun$), and massive galaxies
that may be observable at redshift $\sim$10.  When we combine our data
with simulations that include high mass haloes at low redshift, we find
that the best fit to the halo mass function depends not only on linear
overdensity, as is commonly assumed in analytic models, but also upon
the slope of the linear power spectrum at the scale of the halo
mass. The Press-Schechter model gives a poor fit to the halo mass
function in the simulations at all epochs; the Sheth-Tormen model
gives a better match, but still overpredicts the abundance of rare
objects at all times by up to 50\%.  Finally, we consider the
consequences of the recently released WMAP 3-year cosmological
parameters. These lead to much less structure at high redshift,
reducing the number of $z=10$ ``mini-haloes'' by more than a factor of two and
the number of $z=30$ galaxy hosts by more than four orders of
magnitude. Code to generate our best-fit halo mass function may be
downloaded from
http://icc.dur.ac.uk/Research/PublicDownloads/genmf\_readme.html
\end{abstract}

\section{introduction}

The numbers of haloes in the high redshift universe are critical for
determining the numbers of stars and galaxies at high redshift, for
understanding reionization, and for guiding observational campaigns
designed to search for the first stars and galaxies. The reionization
of the universe is thought to be caused by some combination of
metal-free stars, early galaxies and accreting black holes (see \eg
Bromm \& Larson 2004; Ciardi \& Ferrara 2005; Reed \etal 2005 and
references therein), all of which are expected to lie in dark matter
haloes, the numbers of which are, to date, highly uncertain at these
early times.  This paper presents an array of cosmological simulations
of a wide range of volumes with which we determine the numbers of
haloes over the entire mass range that is expected to host luminous
sources in the high redshift universe.

The first galaxies are expected to form within haloes of sufficiently
high virial temperature to allow efficient cooling by atomic hydrogen
via collisionally induced emission processes, which become strong at
temperatures of $\sim$10$^{4}$K, providing the possibility of
efficient star formation.  Haloes of mass $\sim 10^{8} \times
[10/(1+z)^{3/2}]$ $h^{-1} \msun$ have the required virial temperature 
to host galaxies.  Haloes with virial temperature less than the
threshold for atomic hydrogen line cooling, but larger than
$\sim$2,000K, often referred to as ``mini-haloes'', have the potential
to host metal-free (population III) stars that form from gas cooled
through the production of H$_{2}$ and the resulting
collisionally-excited line emission.  The first stars in the universe
are expected to form within such mini-haloes, which have masses as
small as $\sim 10^{5-6} h^{-1} \msun$ at redshifts of $\sim$10-50.
The inability of collapsing H$_2$-cooled gas to fragment to small
masses, demonstrated in pioneering simulations by Abel, Bryan, \&
Norman (2000, 2002) and by Bromm, Coppi, \& Larson 1999, 2002),
suggests that these first stars will be very massive ($\simgt 100
\msun$), luminous, and short lived, and will thus have dramatic
effects on their surroundings.  These population III stars begin the
process of enriching the universe with heavy elements, and are
expected to have an important impact (directly or indirectly) on
reionization.

Early estimates of the numbers of haloes in the pre-reionized and the
reionizing universe have relied upon analytic arguments such as the
Press \& Schechter (1974) formalism or the later Sheth \& Tormen
(1999; S-T) function 
(further detailed in Sheth, Mo, \& Tormen 2001; Sheth \& Tormen 2002).
For haloes at very high redshifts, which form
from rare fluctuations in the density field, these analytic methods
are in poor agreement with each other. At lower redshifts, halo
numbers have been extensively studied using N-body simulations of
large volumes. Simulations by Jenkins \etal (2001) show that the mass
function of dark matter haloes in the mass range from galaxies to
clusters is reasonably well described by the Sheth \& Tormen (1999)
analytic function out to redshift 5, although with some suppression at
high masses.  Jenkins et al. proposed an analytic fitting formula for
the ``universal" mass function in their simulations. Warren
\etal (2006) used a suite of simulations to measure the redshift zero
mass function to high precision.  Reed \etal (2003) used higher
resolution simulations to show that the broad agreement with the S-T
function persists down to dwarf scales and to $z=10$, a result that
was confirmed by the larger ``Millennium'' simulation of Springel
\etal (2005).  However, at $z \simeq 15$, Reed
\etal (2003) also found fewer haloes than predicted by the S-T
function.  
Qualitatively consistent results 
have been found by Iliev \etal (2006), Zahn \etal (2006). 
These studies indicate that current analytic predictions of
halo numbers are inaccurate at high redshift and demonstrate the need
for N-body studies to determine the mass function at earlier times.

Early attempts to simulate the formation of dark matter haloes in the
young universe suffered from effects resulting from the finite box
sizes of the simulations, as noted by (\eg White \& Springel 2000;
Barkana \& Loeb 2004).  Recently Schneider \etal (2006) have modelled
haloes large enough to host galaxies using {\small PINOCCHIO} (Monaco \etal
2002; Monaco, Theuns, \& Taffoni 2002), a code that predicts mass
merger histories given a linear density fluctuation field.  More
directly, Heitmann \etal (2006) used N-body simulations to show that
haloes large enough to cool via atomic hydrogen transitions, and thus
with the potential to host galaxies at redshifts 10 - 20, are well fit
by the Warren \etal mass function, with the largest haloes suppressed
relative to the S-T function by an amount consistent with that seen in
Reed \etal (2003).

Formation of the first haloes large enough to host galaxies occurs as
early as $z\sim35$ (\eg Gao \etal 2005; Reed \etal 2005), much earlier
than the epochs at which the mass function has been calculated
directly.  The abundance of smaller haloes, capable of hosting
population III stars, but too cool for atomic cooling, remains poorly
constrained by numerical simulations; see however early work by
Jang-Condell \& Hernquist (2001).  The major difficulty is the
computational challenge of performing simulations with very high mass
resolution within a volume that is large enough to sample fully the
cosmological mass perturbation spectrum.  At high redshifts, the
effects of finite box size become particularly important because the
haloes to be sampled represent rare fluctuations in the linear
fluctuation spectrum. Since the mass function is steep, the numbers of
such rare haloes are particularly sensitive to large scale, low
amplitude density fluctuations. Finite box size effects worsen as one
attempts to simulate the smaller volumes needed to resolve lower mass
haloes because fluctuations on the scale of the box become comparable
to those on the scale of the halo.  In $\Lambda$CDM cosmologies with
spectral slope parameter $n_s=1$, the effective, local spectral index
of the perturbation spectrum, $n_{\rm eff}$, approaches $-3$ on the
smallest scales, implying that fluctuations on a broad range of scales
have similar amplitude (see further discussion in \S~2) .  As a
result, proper modelling of the power on scales much larger than the
scale of the halo is important. Simulations of small haloes must
therefore have a large dynamic range in order to model accurately all
of the fluctuations that determine the formation and evolution of a
halo.

Several authors have estimated the effect of the finite simulation box
size on the halo mass function using techniques based on assuming a
simple cutoff in the power spectrum of density fluctuations on scales
larger than the box length (Barkana \& Loeb 2004; Bagla \& Ray 2005;
Power \& Knebe 2006; Bagla \& Prasad 2006).  While these techniques
are able to account for the missing large-scale power, they do not
account for cosmic variance, i.e. the run-to-run variations introduced
by the finite sampling of density modes, particularly at scales near
the box size (\eg Sirko 2005).  Since the density field is derived
from a set a discrete Fourier modes with maximum wavelength equal to
the box size, the power at the largest wavelengths is determined by
only a small number of realised modes.  As a result, each random
realisation of a simulation volume produces different large-scale
structures.  
 
We introduce a technique, described in Section~2, which deals with the
finite volume effects through a mass-conserving transformation of the
halo mass function estimated from each individual simulation output.
In order to verify the ability of this technique to account for finite
volume affects, we perform simulations of a wide, but closely spaced
range of volumes, which results in large overlaps in redshift and in
the range of resolved halo masses in different simulations.  The
agreement of the inferred mass functions in the regimes where halo
masses overlap allows us to verify the finite volume correction, and
also allows us to rule out resolution dependencies of our results.
Multiple realisations of a single volume at identical resolution then
test how well the correction to the inferred mass function is able to
minimise the effects of cosmic variance. 

Our simulations are designed to extend the mass function to small
masses and high redshifts, covering a mass range of $10^{5}$ to
10$^{12}$ $h^{-1} \msun$, at redshifts 10 to 30, and we supplement
them with low redshift data taken from other studies.  This extends
the mass function down to masses small enough to include the
``mini-haloes'' capable only of hosting stars formed via
H$_{2}$-cooling, and determines more precisely the mass function of
larger haloes which can host galaxies.  In \S~2, we define the halo
mass function and outline our method for dealing with finite volume
effects. In \S~3, we discuss our suite of simulations of varying box
sizes and resolutions.  In \S~4, we demonstrate the effectiveness of
our techniques for correcting for finite volume and cosmic variance.
We then present our mass function and compare it to previous works.
In \S~5, we consider the dependence of the mass function on
cosmological parameters in the light of the recent WMAP third year
results (Spergel \etal 2006).  In \S~6, we discuss some implications
of our mass function for astrophysical models that rely on the mass
function of high redshift haloes.  Finally, our conclusions are
summarised in \S~7.

Except when otherwise indicated, we assume throughout a flat
$\Lambda$CDM model with the following cosmological parameters, which
are consistent with the combined first year WMAP/2dFGRS results
(Spergel \etal 2003): matter density, $\Omega_m=0.25$; dark energy
density, $\Omega_\Lambda=0.75$; baryon density, $\Omega_{\rm
baryon}=0.045$; fluctuation amplitude, $\sigma_{\rm 8}=0.9$; Hubble
constant $h=0.73$ (in units of 100 km s$^{-1}$ Mpc$^{-1}$); and no
tilt (i.e. a primordial spectral index of 1).  Note that our results
should, in principle, be scalable to other values of cosmological
parameters.

\section{The halo mass function}

In this section we define the notation that we use to describe the
halo mass function and introduce our method for estimating the halo
mass function from our N-body simulations.  The simulations themselves
are described in \S~3.
 
\subsection{Definitions}

The differential halo mass function, or halo mass function for short,
${\rm d}n/{\rm d}M$, is defined as the number of haloes of mass $M$ per
unit volume per unit interval in $M$.  In this section, we introduce,
for reasons that will become apparent later, an alternate pair of
variables, $f(\sigma)$ and $\ln\sigma^{-1}$, to describe the halo mass
function. The quantity $\sigma$ is the RMS linear overdensity of the
density field smoothed with a top-hat filter with a radius that
encloses a mass $M$ at the mean cosmic matter density.  For an
infinite volume we have:

\begin{equation}
\sigma^2(M)  =  {b^{2}(z)\over2\pi^2}\int_0^\infty
k^2P(k)W^2(k;M){\rm d}k,
\label{varinf}
\end{equation}
where $P(k)$ is the linear power spectrum of the density fluctuations
at $z=0$, $W(k;M)$ is the Fourier transform of the real-space top-hat
filter, and ${\it b(z)}$ is the growth factor of linear perturbations
normalised to unity at $z=0$ (Peebles 1993).  The quantity
$\ln\sigma^{-1}$ can be thought of as a mass variable in the sense
that higher values of $\ln\sigma^{-1}$ correspond to higher masses for
a given redshift and matter power spectrum.

The quantity, $f(\sigma)$, to which we will refer as the mass
function, is defined as the fraction of mass in collapsed haloes per
unit interval in $\ln\sigma^{-1}$.  If all matter is in haloes of some
mass then: 
\begin{equation}
  \int_{-\infty}^\infty f {\rm~d}\ln\sigma^{-1} = 1.
\label{integcond}
\end{equation}
  The differential halo mass function is related to $f(\sigma)$ by:
\begin{equation}
{{\rm d}n\over{\rm d}M}  =  {\rho_0\over M}{{\rm d}\ln\sigma^{-1}\over{\rm d}M}
f(\sigma),
\label{diffrel}
\end{equation}
where $\rho_0$ is the mean mass density of the universe.

The function $f(\sigma)$ will depend on how haloes are defined.  For
this paper, in common with most recent work on halo mass functions, we
use the friends-of-friends (FOF) algorithm (Davis \etal 1985) with a linking
length of 0.2 times the mean inter-particle separation.  In the
appendix we include mass functions using the SO algorithm (Lacey \& Cole
1994).

The reasons for choosing to describe the halo mass function in terms
of the rather abstract variables $f$ and $\ln\sigma^{-1}$ are
twofold. Firstly, the most commonly used analytical halo mass functions
can be expressed compactly in terms of these variables.  For example, 
the Press \& Schechter (1974; P-S) mass function can be expressed as:
\begin{equation}\label{ps_massfn}
   f_{\rm P-S}(\sigma) = \sqrt{2\over\pi}
   {\delta_c\over\sigma}\exp\bigg[-{\delta_c^2\over2\sigma^2}\bigg],
\end{equation}
where the parameter $\delta_c=1.686$ can be interpreted physically
as the linearly extrapolated overdensity of a top-hat spherical density
perturbation at the moment of maximum compression 
for an Einstein de-Sitter universe ($\Omega_{m}=1$).  
The evolution of $\delta_c$,
predicted by the 
spherical collapse model (\eg Eke, Cole \& Frenk 1996)
as $\Omega_{m,z}$ transitions from $\simeq 1$ at high 
redshift to its present value,
is sufficiently weak that we have ignored it in our treatment.
Similarly, the Sheth-Tormen (S-T) mass function takes the form:

\begin{equation}\label{stfunc}
  f_{\rm S-T}(\sigma) = A\sqrt{{2a\over\pi}}
\bigg[1+\big({\sigma^2\over a\delta_c^2}\big)^p\bigg]
{\delta_c\over\sigma}\exp\bigg[-{a\delta_c^2\over2\sigma^2}\bigg].
\end{equation} 
The choice of values $A=0.3222$, $a=0.707$ and $p=0.3$ provides a
significantly better fit to mass functions determined from numerical
simulations over a wide range of masses and redshifts than the P-S
formula.

Secondly, it has been found empirically, consistently with the analytic
mass function formulae above, that the FOF mass function determined
from cosmological simulations for a wide range of redshifts, and for a
wide range of cosmological models can be fitted accurately by a unique
function $f(\sigma)$ (e.g. S-T 1999, 
Jenkins \etal 2001, Reed \etal 2003, Linder
\& Jenkins 2003, Lokas, Bode \& Hoffmann 2004, Warren \etal 2006). A
number of formulae for $f(\sigma)$ have been proposed based on fits to
simulation data and these are generally consistent at the
$\sim$10-30\% with the largest differences occuring at the high mass
end, where the rarity and steepness of the halo mass function make its
estimation rather challenging.  The main aim of this paper is to
determine the halo mass function at high redshift and to provide a
fitting formula which applies to both our high and low redshift
simulation data.

It is appropriate here to question whether the halo mass function can
really be expressed as a universal function of the form $f(\sigma)$
(see further discussion in S-T 1999).
Structure formation in Einstein-deSitter cosmological N-body
simulations in which the initial power spectrum is a (truncated)
power-law ($P(k)\propto k^n$) has been found to show self-similar
evolution (e.g. Efstathiou \etal 1998, Lacey \& Cole 1994).  For a
given value of $n$, self-similar evolution implies a universal form
for $f(\sigma)$.  However the function $f$ could, in principle, be
different for different values of the power-law index, $n$.  There is
suggestive evidence that this may indeed be the case in Fig.~1 of Lacey \& Cole (1994).

Thus, while empirically the CDM halo mass function appears to be well
described by a function $f(\sigma)$, it may be that it is possible to
improve the accuracy of a fitting formula 
by adding an additional parameter.  For the CDM power spectrum where
the slope curves gently a natural parameter to take would be the local
slope of the power spectrum.  At the small spatial scales 
relevant to the halo mass function at high redshift, the
spectral slope approaches a critical value $n=-3$ which marks the boundary
between bottom-up hierarchical structure formation and top-down
structure formation.  One might expect that the need for an extra
parameter would become apparent as this boundary is approached.  In
\S~4.2, we find that we can improve the goodness of fit to our
mass function by using this as an extra parameter, and it is the high
redshift simulation data which require this.
   
\subsection{Finite simulations and the global mass function}

Due to limited computing resources any cosmological N-body simulation
can only model a finite volume of space. Periodic boundary conditions
are usually implemented in order to avoid edge effects, with the most
common geometry being a periodic cube.  In this case, the overdensity
of matter, $\delta({\bf r})$, is given by a sum over Fourier modes:
\begin{equation}
\delta  =  \sum_{\bf k}\delta_{\bf k} \exp(i{\bf k}.{\bf r}),
\label{deltap}
\end{equation}
   where the $\delta_{\bf k}$ are complex amplitudes which obey a
reality condition $\delta_{\bf k}^* = \delta_{\bf -k}$.  Because the
simulation volume must have mean density, $\delta_{\bf k=0}=0$. The
Fourier modes have wavenumbers ${\bf k} = 2\pi/L_{\rm box}(l,m,n)$,
where $l,m,n$ are integers and $L_{\rm box}$ is the side-length of the
simulation volume.  

The initial conditions for a simulation of a CDM universe
with adiabatic density perturbations require that the initial density
field should be a Gaussian random field. In this case, the phases of the wave
amplitudes, $\delta_{\bf k}$, are random, and the amplitude of each mode
is drawn from a Rayleigh distribution (Efstathiou \etal 1985) where:
\begin{equation}
<|\delta_{\bf k}|^2> = P(k)/L_{\rm box}^3
\label{gaus_ampl}
\end{equation}
 and the brackets $<>$ denote an ensemble average over realisations.

For a periodic cosmological simulation, the smoothed rms linear
overdensity, $\sigma$, is given by the discrete analog of
Eqn.~\ref{varinf}:

\begin{equation}
\sigma^2(M)  =  {b^2}(z)\sum_{\bf k}|\delta_{\bf k}|^2 W^2(k;M),
\label{varsim}
\end{equation}
where $|\delta_{\bf k}|$ refers to the linear amplitude of the Fourier
modes at $z=0$, and $b$ and $W$ are the same as in Eqn.~\ref{varinf}.

A number of authors (Power \& Knebe 2006; Sirko 2005; Bagla \& Prasad
2006) have highlighted the problem that the halo mass function in a
finite periodic box will differ between realisations and that the
ensemble average mass function will not be the same as in the limit of
an infinitely large simulation volume. The effects of having a
discrete power spectrum with only a small number of modes with
wavelengths comparable to the size of the cube and no power with
wavelengths larger than the cube are particularly important for small
cosmological volumes where the contribution to the variance of the
density field from each successive decade of wavenumber is a very
weakly increasing function. Given the computing resources available to
us and the requirement that we should resolve haloes with a hundred or
more particles at high redshift, it is inevitable that the volumes we
wish to simulate will be affected significantly by finite box effects.

Our approach to minimise finite box effects and to estimate the high
redshift halo mass function is to make the ansatz that the universal
form of the halo mass function correctly describes the halo mass
function even in volumes where finite volume effects are
significant. For small volumes it is important to use the correct
relation between $M$ and $\sigma$ given by Eqn.~\ref{varsim} for {\it
each individual simulation} in order to estimate the halo mass
function in $f(\sigma)-\ln\sigma^{-1}$ space. Having estimated
$f(\sigma)$ we can now predict the halo mass function, Eqn.~\ref{diffrel} for the
astrophysically interesting case of an infinite volume, using the
relation between $\sigma$ and $M$ given by Eqn.~\ref{varinf}. We call
this mass function the global mass function.

It not obvious a priori just how successful this approach will be in
recovering the global mass function.  
However, it is worth noting that this approach contains 
inherently the
essential elements present in conditional mass function methodology
wherein the number of 
haloes within a local patch is estimated
(Mo \& White 1996, Bower 1991, Bond \etal 1991, Lacey \& Cole 1993).
If we take our simulation volume to be a patch of universe
with mass $M_{\rm patch}$ and mean density, 
then the variance that we measure within the simulation volume is close to 
$\sigma^2_{\rm global}(M)-\sigma^2_{\rm global}(M_{\rm patch})$, 
where $\sigma^2_{\rm global}(M)$ is given by Eqn.~\ref{varinf}.
The mass function in such a finite patch is reliably predicted by substituting 
$\sigma^2_{\rm global}(M)-\sigma^2_{\rm global}(M_{\rm patch})$ for 
$\rm \sigma^2_{\rm global}(M)$ into the functional form of the mass function
(Sheth \& Tormen 2002). 
Our methodology is an improvement on such an approach because we also include 
the effects of run to run ``cosmic'' variance.
As will be discussed in
\S~4, we find that our method does work very well in practice.
To demonstrate this one needs a large number of simulations with
multiple random realisations at a fixed box size and a variety of
differing box sizes.  We describe our suite of simulations in the next
Section.

In practice, we determine $\sigma(M)$ for a particular simulation by
measuring the power spectrum of the initial conditions.  We perform a
sum over the low-k modes, but switch to an integral over the linear
power spectrum for wavenumbers greater than 1/20th of the particle
Nyquist frequency of the simulation. At the changeover point the
number of independent modes is sufficiently large that the difference
between doing a summation or an integration is negligible.

Our procedure for correcting for the finite simulation volume
is more direct than and is simpler in practice than the Cole (1997)
modification of the Tormen \& Bertschinger (1996) mode adding
procedure (MAP).  In the MAP algorithm, an evolved simulation is
replicated onto a linear density field of a larger volume, adding
displacements from the long wavelengths to the replicated particle
positions to approximate the effects of large scale power, thereby
increasing the effective volume of the simulation.  Adding a large
scale density perturbation has the effect of changing the local value
of $\Omega_{\rm m}$, which also changes the linear growth factor, and
in effect, changes the redshift.  This means that in order to produce
the equivalent of a large volume simulation snapshot, replicated
particles from the small volume simulation must be temporally
synchronized according to what large scale density they are being
``mapped'' onto (Cole 1997).  Although the MAP approach is promising,
it has not been thoroughly tested in the regime in which we are
interested, and it does not take into account directly the coupling
between small and large scales, which has the potential to affect halo
formation.  

\section{The simulations} 

We use the parallel gravity solver
{\small L-GADGET2} (Springel \etal 2005) to follow the evolution of dark matter
in a number of realisations of different cosmological volumes.  Table
\ref{simtable} lists all our simulations and the numerical parameters
used.  The highest resolution simulations have particle mass of
10$^{3}$ $h^{-1} \msun$ and resolve haloes to redshifts as high as 30.
Our new simulation volumes range from 1 to 100 $h^{-1}$Mpc on a side.
For these runs, the cell size of the mesh used by {\small L-GADGET2} in the PM
portion of the Tree-PM force algorithm to compute long range
gravitational forces is equal to one half the mean particle spacing.
We also include results of the 500 $h^{-1}$Mpc ``Millennium run''
(Springel \etal 2005).  For low redshift haloes, we include analysis
of the 1340 $h^{-1}$Mpc run by Angulo \etal (2006, in preparation),
and the 3 $h^{-1}$Gpc ``Hubble volume'', which has $\Omega_m=0.3$,
$\Omega_\Lambda=0.7$, and $\sigma_{\rm 8}=0.9$, was run using HYDRA
(Couchman, Thomas \& Pearce 1995; Pearce \& Couchman 1997), and uses
the Bond \& Efstathiou (1984) transfer function (see Colberg \etal
2000 and Jenkins \etal 2001 for details).  We have verified the
robustness of our results to the choice of run parameters by varying
individually the starting redshift (${\rm z_{start}}$), the fractional
force accuracy (${\rm \Delta_{force~acc}=0.005}$), the softening
length (${\rm r_{soft}}$), and the maximum allowed timestep
($\Delta_t\equiv\Delta\ln(1+z)^{-1}$); these tests are detailed in the
Appendix.

Initial conditions for runs with box length of 50 $h^{-1}$Mpc or
smaller were created using the CMBFAST transfer function (Seljak \&
Zaldarriaga 1996) as follows.  Traditionally, the initial conditions
are generated from a transfer function calculated at $z=0$ and
extrapolated to the initial redshift using linear theory.  However, in
order to avoid a high wavenumber ($k$) feature\footnote{ We noticed a
feature in the CMBFAST transfer function at
$k\sim10^{2.5}~h$Mpc$^{-1}$, where the slope steepens for
approximately a decade in $k$, resulting in a power spectrum that
briefly becomes steeper than the theoretical asymptotic minimum slope
of $n=-3$ at the smallest scales for a primordial spectral index
$n_s=1$.  This unexpected feature is not present in high redshift ($z
\gg 100$) computations of the CMBFAST transfer function.}  in the
CMBFAST $z=0$ transfer function, our adopted transfer function
consists of the $z=0$ transfer function for small $k$ spliced together
at $k = 1 ~h$Mpc$^{-1}$ with a high redshift ($z=599$) transfer
function for large $k$.  Power is matched on either side of the
splice.  The location of the splice is chosen to be at a point where
the shape of the transfer function has essentially no redshift
dependence, thereby ensuring continuity of spectral slope.  We use a
combined mass-weighted dark matter plus baryon transfer function.  
The
100 $h^{-1}$Mpc run and the 1340 $h^{-1}$Mpc runs both used the
transfer function used for the Millennium simulation, which is
detailed in Springel \etal (2005).

Our choice of initial conditions and simulation techniques
neglects direct treatment of baryons.  The method that we implement is
ideal for our purposes of modelling the dark matter halo mass function
and assessing its universality given an input dark matter fluctuation
spectrum.  However, for the purpose of making highly accurate
predictions of the numbers of haloes in the real universe, coupling of
baryons to photons, and subsequently to dark matter can be important
at the high redshifts that are involved.  For example, at the starting
redshift, the baryons are much more smoothly distributed than the dark
matter.  The ensuing evolution of the dark matter distribution is then
affected as the baryon fluctuations begin to catch up to the dark
matter.  We refer the reader to further discussions regarding these
and related issues by \eg Yamamota, Noashi \& Sato (1998), Yoshida,
Sugiyama \& Hernquist (2003), Naoz \& Barkana (2005) and Naoz, Noter
\& Barkana (2006).

\begin{table}
\caption{Simulation parameters. N$_{\rm runs}$ random realisations of
cubical volumes of side L$_{\rm box}$ were simulated, from redshift
$z_{\rm start}$ to $z_{\rm fin}$.  N$_{part}$ particles of mass
M$_{\rm part}$ and gravitational force softening length r$_{\rm soft}$
were employed.}

\begin{tabular*}{\columnwidth}{@{}lllllll}
\hline\hline
 N$_{\rm runs}$   &  L$_{\rm box}$  & M$_{\rm part}$  & N$_{\rm part}$
& z$_{\rm fin}$ & z$_{\rm start}$ & r$_{\rm soft}$ \\ 
    &  $h^{-1}$ Mpc & $h^{-1} \msun$ &    &    &     & $h^{-1}$kpc \\
\hline
11 & 1.0 & 1.1 $\times$ 10$^{3}$ & 400$^{3}$ & 10 & 299 & 0.125 \\
1 & 2.5 & 1.1 $\times$ 10$^{3}$ & 1000$^{3}$ & 10 & 299 & 0.125 \\
3 & 2.5 & 1.1 $\times$ 10$^{3}$ & 1000$^{3}$ & 30 & 299 & 0.125 \\
1 & 2.5 & 8.7 $\times$ 10$^{3}$ & 500$^{3}$ & 10 & 299 & 0.25 \\
1 & 2.5 & 1.4 $\times$ 10$^{5}$ & 200$^{3}$ & 10 & 299 & 0.625 \\
1 & 4.64 & 1.1 $\times$ 10$^{5}$ & 400$^{3}$ & 10 & 249 & 0.58 \\
2 & 11.6 & 1.1 $\times$ 10$^{5}$ & 1000$^{3}$ & 10 & 249 & 0.58 \\
1 & 20 & 8.7 $\times$ 10$^{6}$ & 400$^{3}$ & 10 & 249 & 2.5 \\
2 & 50 & 8.7 $\times$ 10$^{6}$ & 1000$^{3}$ & 10 & 299 & 2.4 \\
1 & 100 & 9.5 $\times$ 10$^{7}$ & 900$^{3}$ & 10 & 149 & 2.4 \\
1 & 500$^{\dagger}$ & 8.6 $\times$ 10$^{8}$ & 2160$^{3}$ & 0 & 127 & 5.0 \\
1 & 1340$^{\dagger\dagger}$ & 5.5 $\times$ 10$^{10}$ & 1448$^{3}$ & 0 & 63 & 20 \\
1 & 3000$^{\dagger\dagger\dagger}$ &  2.2 $\times$ 10$^{12}$ & 1000$^{3}$ & 0 & 35 & 100 \\
\hline

\label{simtable}
\end{tabular*}
$^{\dagger}$ ``Millennium'' run (Springel \etal 2005) \\
$^{\dagger\dagger}$ Angulo \etal (2006)\\
$^{\dagger\dagger\dagger}$ ``Hubble Volume'' (Colberg \etal 2000; Jenkins \etal 2001)\\
\end{table}

\section{results}

\subsection{The mass function}
\begin{figure*}
  \begin{center}
 
    \epsfig{file=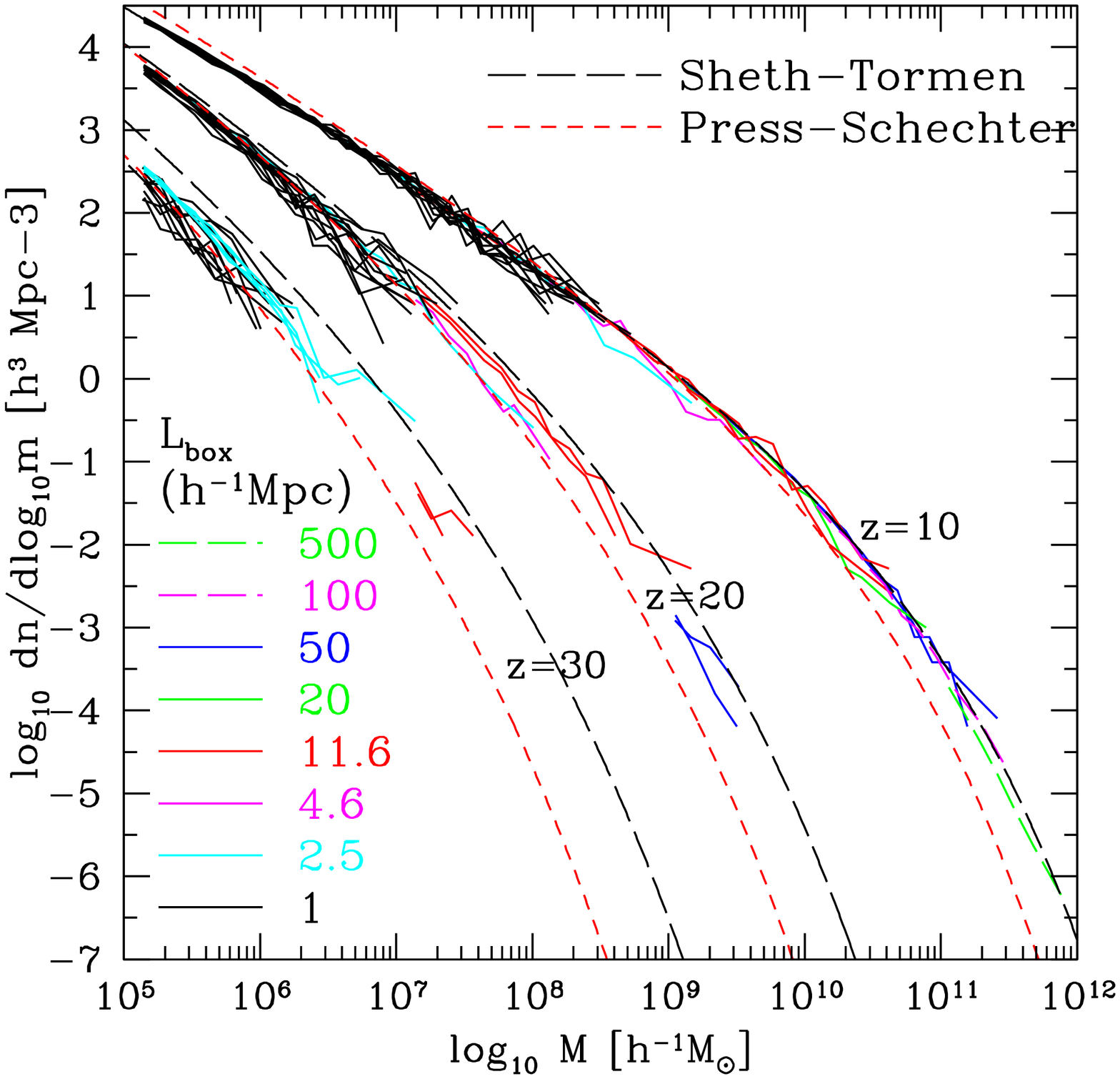, width=0.49\textwidth}
    \epsfig{file=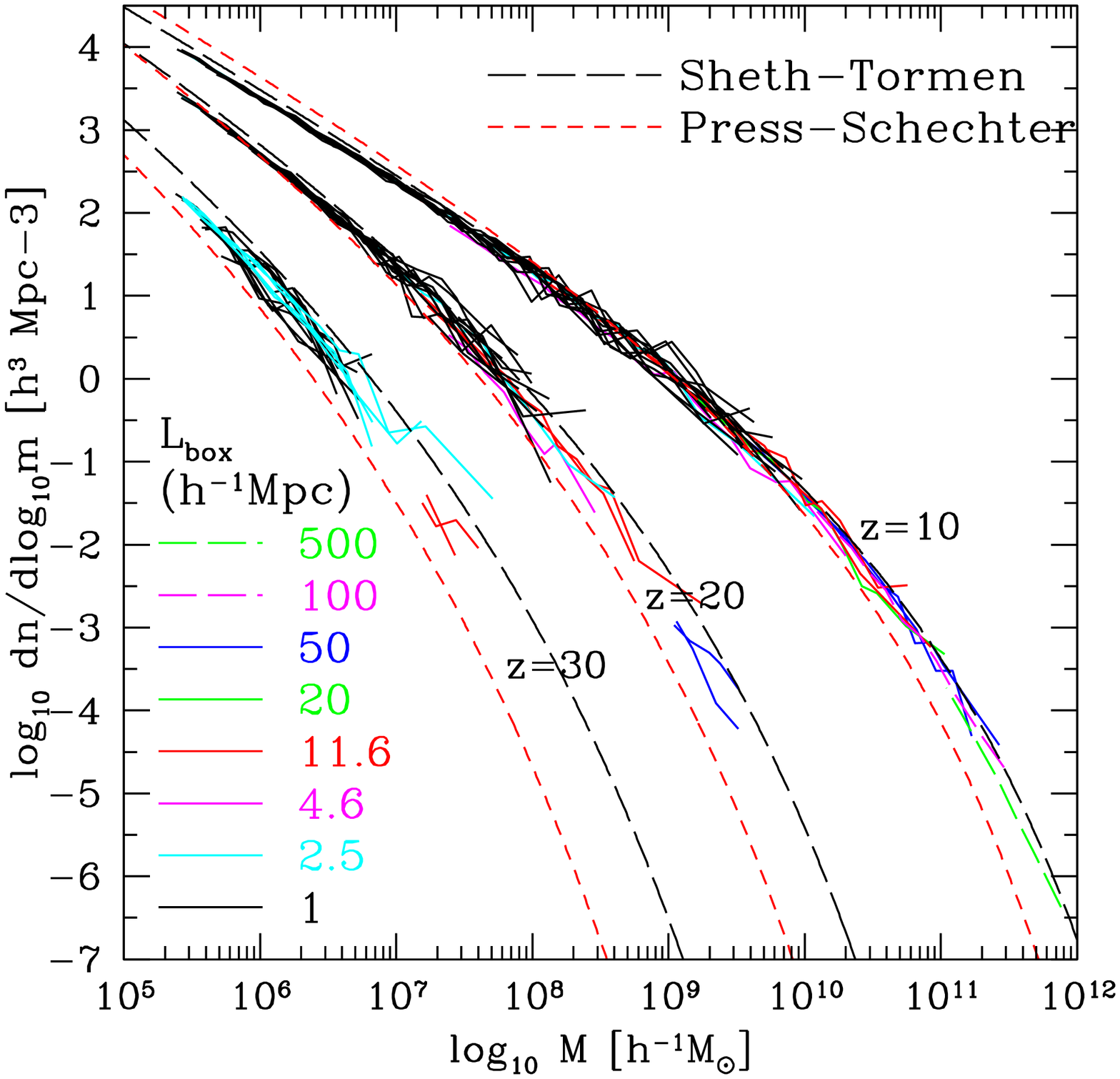, width=0.49\textwidth}
 
$\begin{array}{c@{\hspace{0.2\textwidth}}c}
\mbox{\bf (a) Raw simulation mass function} & \mbox{\bf (b) Global mass function}
\end{array}$
 
\caption{Differential simulated mass function of {\it
friends-of-friends} dark matter haloes for redshift 10, 20, and 30
compared with the Sheth \& Tormen and Press \& Schechter analytic
predictions.  The corrected mass function (right panel) makes a
correction for cosmic variance and for finite volume to the simulation
mass fluctuation spectrum by using the relation between ${\rm
\sigma^2}$ and mass derived from the power spectrum of the initial
particle distribution for each realisation; \ie the left panel uses
${\rm \sigma_{\infty}^2(m)}$ (Eqn. \ref{varinf}, the variance versus
mass for an infinite universe) and the right panel utilises ${\rm
\sigma_{sim}^2(m)}$ (Eqn. \ref{varsim}, the actual variance for each
realisation).  Note the reduced run to run scatter and increased
amplitude of the corrected mass functions for small boxes.  For
comparison purposes, the 100 and 500 $h^{-1}$Mpc runs have been
rescaled by the ratio of their expected S-T mass functions to account
for their mildly different transfer functions.  The bin width here and
throughout the paper is $dlog_{10}M=0.125$. }
\label{dnplot}
\end{center}
\end{figure*}

In Fig. \ref{dnplot}, we show the simulation mass functions at
redshifts ten, twenty, and thirty.  The left panel shows the measured
raw abundance of haloes within the (finite) simulation volumes.  
In the right panel, the global mass function is plotted using the
transformation explained in section~2.2.  Operationally this is done
as follows. The group finder returns a group catalogue for a
simulation consisting of a list of the number of haloes of each
mass. Suppose in the catalogue there are $J$ haloes 
with an average mass $M$.
Eqn.~\ref{varsim} is used to find the value of $\sigma$ that corresponds with mass $M$ for this particular simulation.  
Applying Eqn.~\ref{varinf} we can
determine a mass $M^\prime$ which, for an infinite volume, has this
same value of $\sigma$. We can effectively `correct' the catalogue to
yield a new catalogue for the same volume of space as the original
simulation but sampled from an infinitely large simulation volume. To
do this, each of the masses $M$ in the catalogue is replaced with the
corresponding mass $M^\prime$, and the number of haloes $J$ is replaced
by a value $J^\prime$, such that for mass to be conserved in the
transformation $JM= J^\prime M^\prime$.  Note that while $J$ is an
integer, $J^\prime$ will not, in general, also be one. The corrected
catalogue can then be used to construct an estimate of the global
differential halo mass function.

Because of the missing power in smaller volumes, the net effect of the
transformation is an increase in the mass and a decrease in the
abundance of a given bin in such a way that the resulting adjusted
mass function is higher at a given mass.  Note that, with the
correction, the variation between simulations in Fig.~\ref{dnplot} is
very much reduced and the agreement between different box sizes is
much improved.  Once this transformation has been made, the simulated
mass function lies nearer, but generally below the Sheth-Tormen
function for the most massive objects at redshift ten to thirty.  The
Press-Schechter mass function is a poor match to the simulation data,
especially at high masses.

\begin{figure}
\epsfig{file=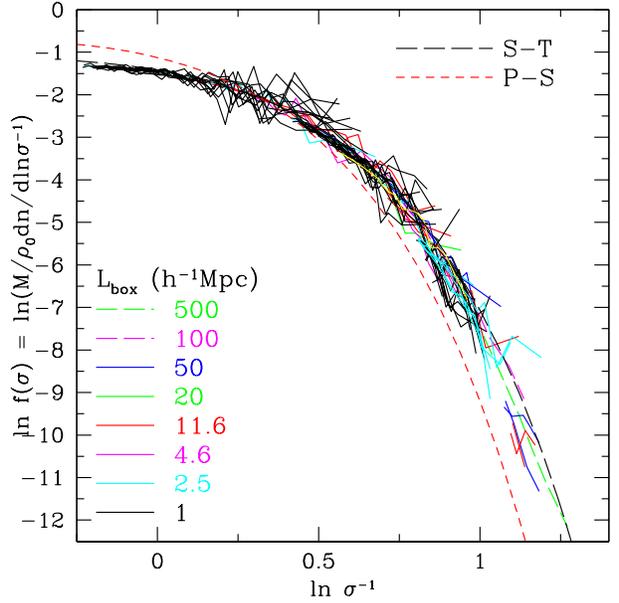, width=\hsize}
\caption{The fraction, f($\sigma$), of collapsed mass per unit $\ln
\sigma^{-1}$, where $\sigma^2$ is the variance, at $z=10$, 20, and 30.
The mass fraction has been adjusted to account for finite volume
effects as described in \S~2.2.  Approximate redshift invariance is
indicated by the fact that all redshifts have roughly the same mass
function.  }
\label{fsigma}
\end{figure}

It is instructive to plot the fraction of collapsed mass, $f(\sigma)$,
as a function of $\ln\sigma^{-1}$.  This fraction is independent of
redshift according to the principles underlying P-S or S-T models.  In
Fig.\ref{fsigma}, we plot $f(\sigma)$ as a function of $\ln
\sigma^{-1}$, including the correction for finite volume.  The fact
that the data over a wide range of redshifts all coincide
approximately in a single form is an indication of the general
redshift independence of the mass function.  However, we discuss in
\S~3.2 some evidence for a weak dependence on redshift.  Haloes formed
from rare fluctuations -- high mass, high redshift, or both -- lie at
large values of $\delta_c/\sigma$ and hence large $\ln \sigma^{-1}$.
Here the mass function is steepest.  Note that rarer haloes do not
necessarily have lower spatial abundance. This can be understood by
comparing a high redshift low mass halo with a low redshift high mass
halo, each forming from $\eg$ a 5-$\sigma$ fluctuation
([$\delta_c/\sigma(M,z)]=5$).  In the case of the low mass, high
redshift halo, the number of regions per comoving volume element that
contain the halo's mass is larger, which results in a higher comoving
halo abundance.

\begin{figure*}
  \begin{center}
 
    \epsfig{file=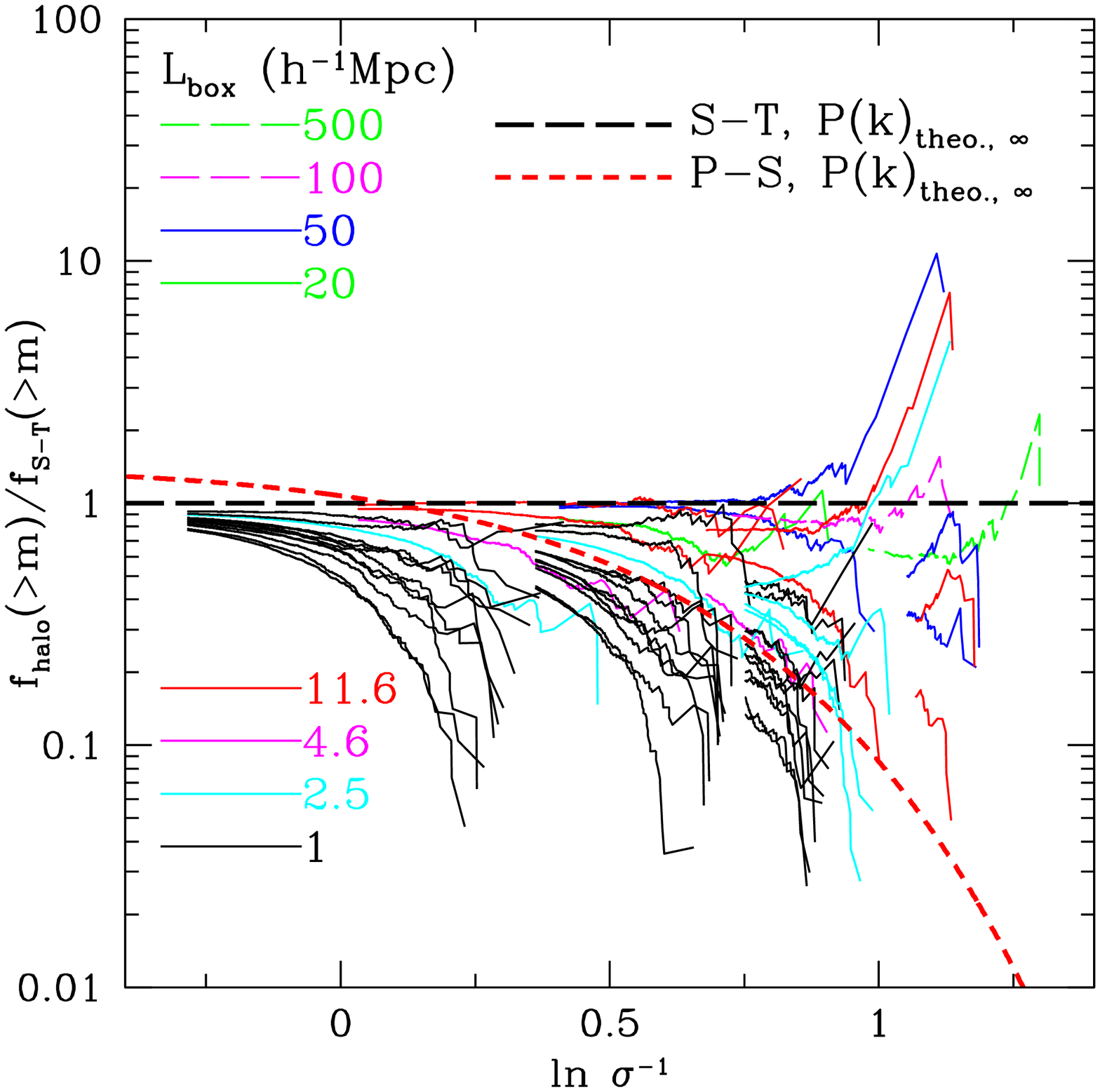, width=0.49\textwidth}
    \epsfig{file=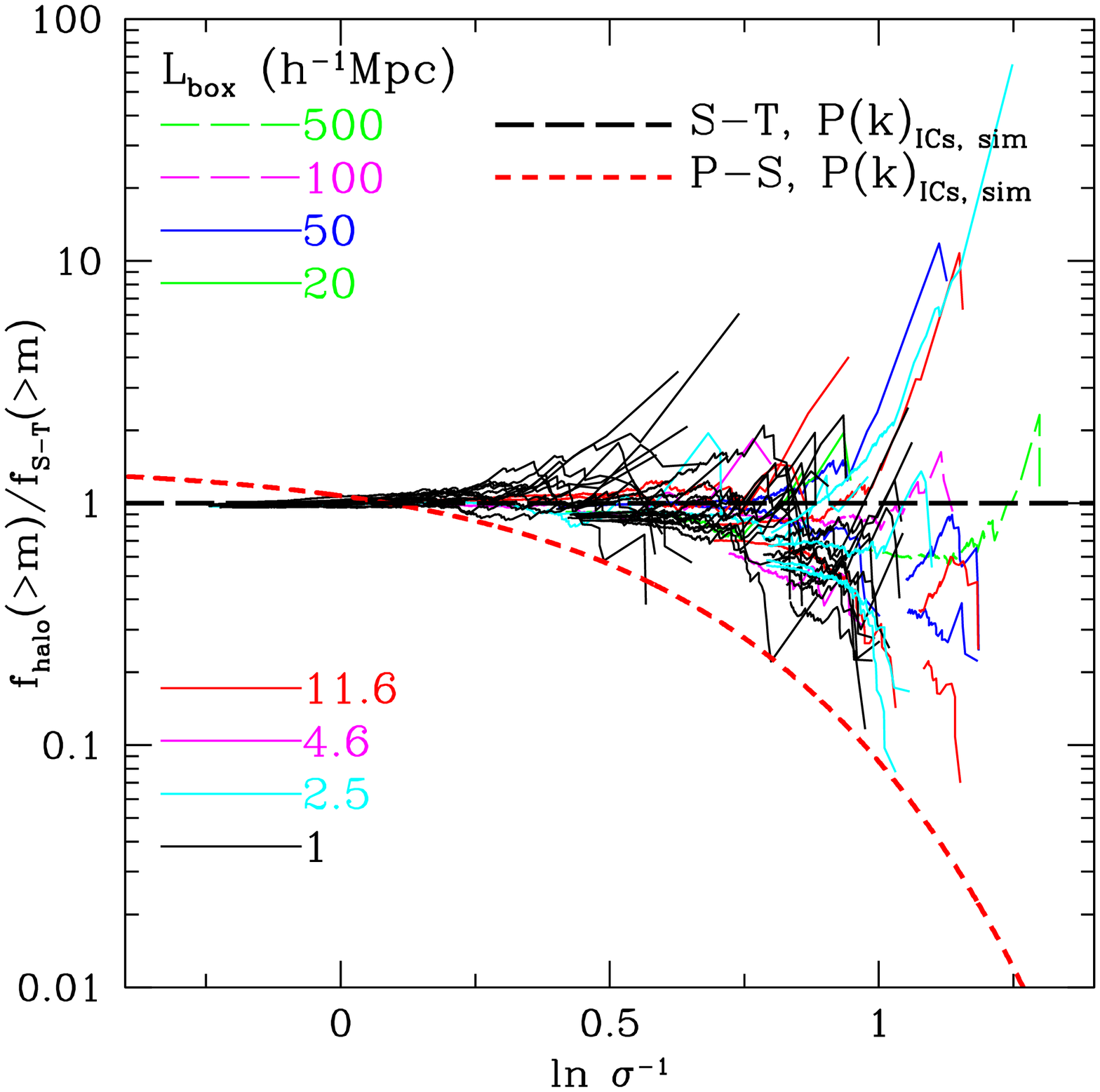, width=0.49\textwidth}
 
$\begin{array}{c@{\hspace{0.2\textwidth}}c}
\mbox{\bf (a) Raw simulation mass function} & \mbox{\bf (b) Global mass function}
\end{array}$
 
\caption{Cumulative fraction of mass contained in FOF haloes. Left
panel assumes ${\rm \sigma_{\infty}(m)}$ (Eqn.. \ref{varinf}), the
variance for an infinite universe).  Right panel utilises ${\rm
\sigma_{sim}(m)}$ (Eqn. \ref{varsim}), the variance-mass relation
derived from the mass power spectrum of the initial particle
distribution.  Accounting for missing large-scale power in this way
results in the systematic upward shift in the corrected mass function.
Reduced run-to-run scatter and improved agreement between different
box sizes (different colours) indicates the effectiveness of the finite
volume correction to the mass function.}
\label{fgtmplot}
\end{center}
\end{figure*}

An important quantity is the cumulative fraction of mass contained in
haloes. In Fig. \ref{fgtmplot}, we plot the ratio of $f(>M)$ divided
by the S-T function.  In the right panel, the global mass function has
been corrected using the actual power from each simulation in the
relation between halo mass and variance (Eqn. \ref{varsim}), as in
Fig. \ref{dnplot}, which automatically accounts for finite volume
effects.  The greatly reduced run-to-run scatter compared to the raw,
uncorrected mass function highlights the improvement gained by using a
more accurate relation between halo mass and variance.  The correction
for limited simulation volume is evidenced by the systematic upward
shift in the cumulative mass fraction, which is strongest for small
boxes, and for high mass and/or redshift ``rare'' haloes.

\subsection{Fitting the mass function}

\begin{figure}
\epsfig{file=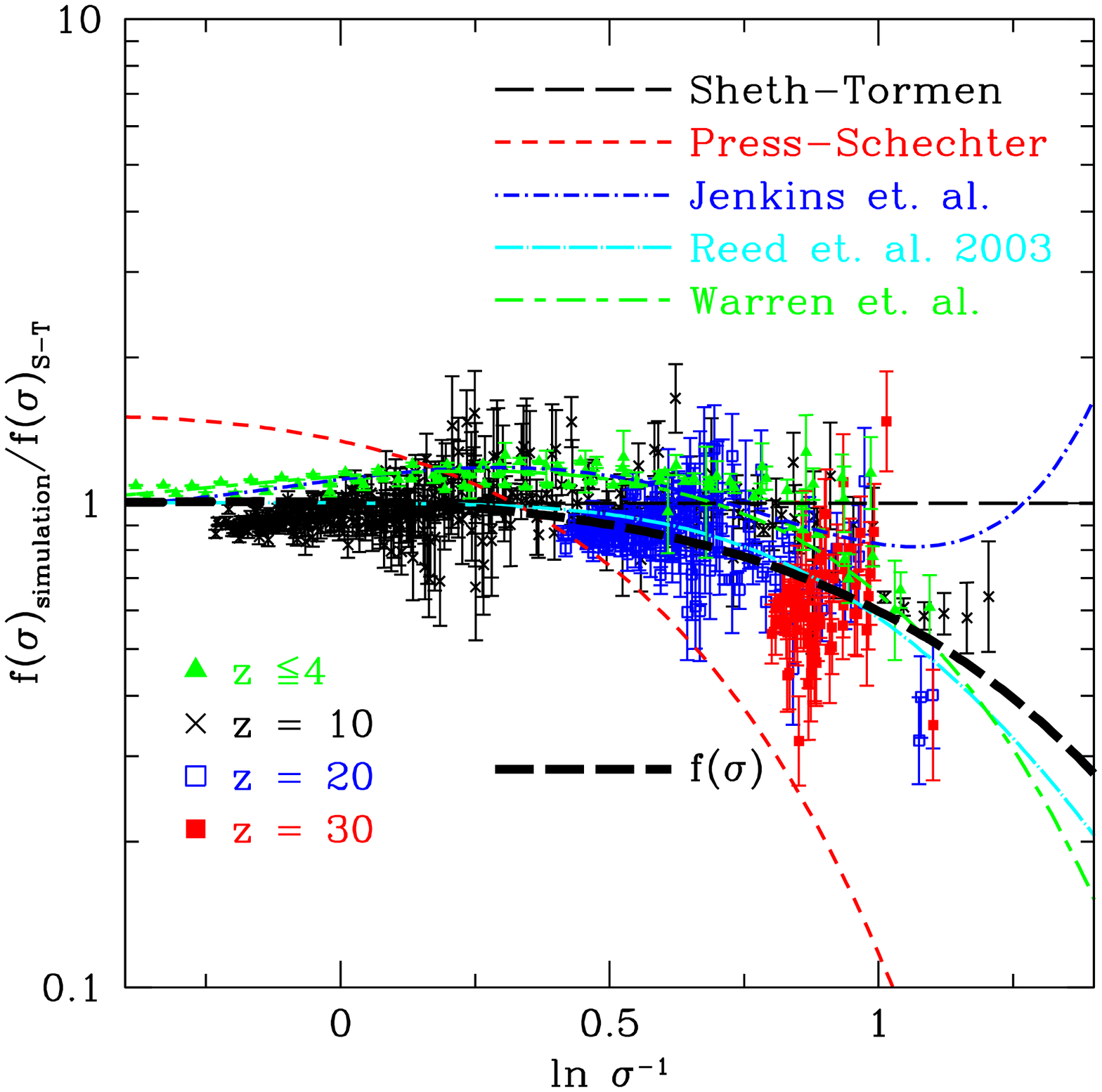, width=\hsize}
\caption{Differential global (corrected for finite volume; see text)
mass function of {\it friends-of-friends} dark matter haloes for
redshift 0, 1, 4, 10, 20, and 30 compared with analytic fits. Thick
curves correspond to the modified S-T function
(Eq. \ref{stmod1param}).  
Error bars denote 1-$\sigma$ poisson uncertainties.
}
\label{1fit}
\end{figure}

\begin{figure*}
  \begin{center}
 
    \epsfig{file=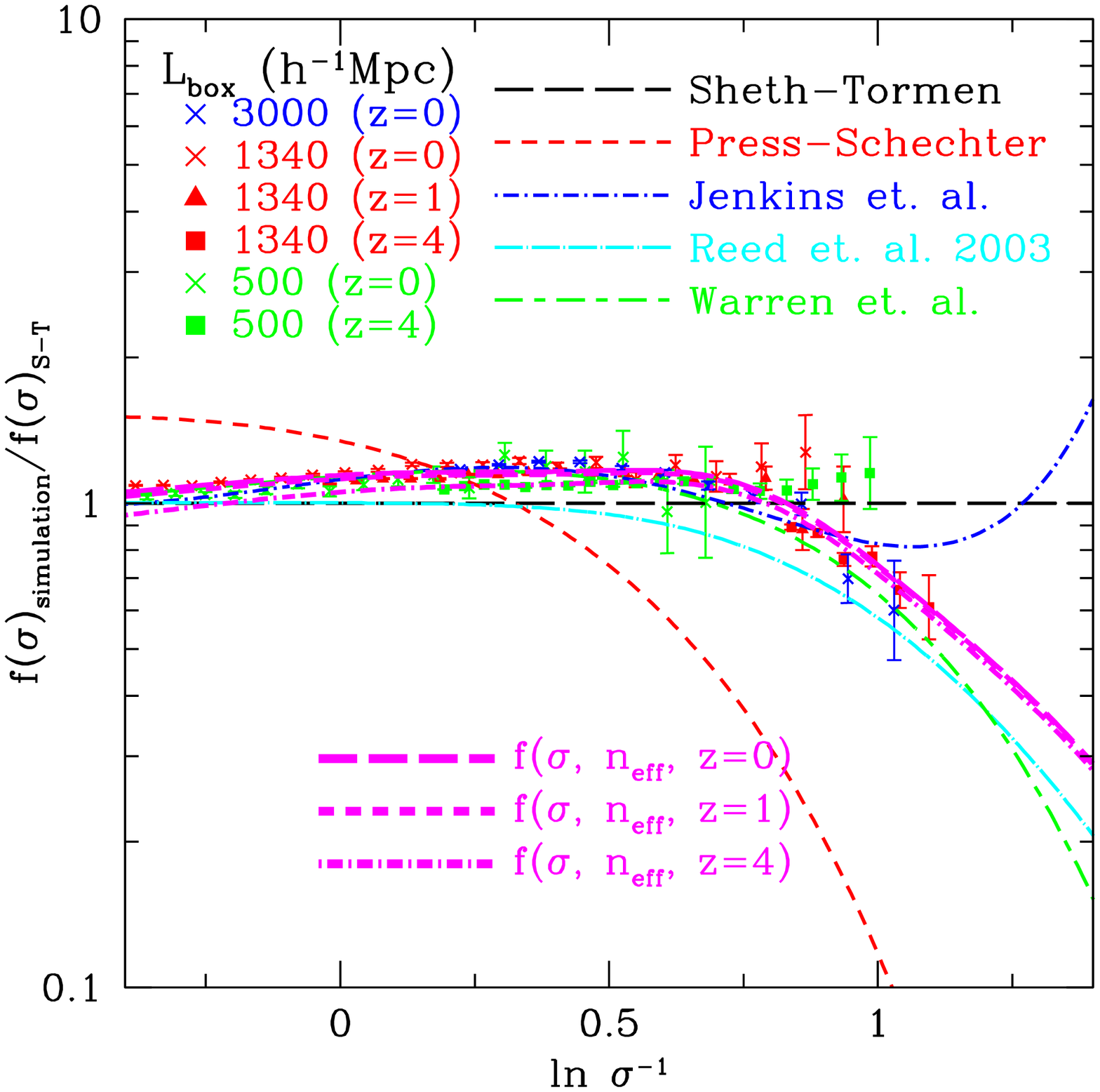, width=0.49\textwidth}
   \epsfig{file=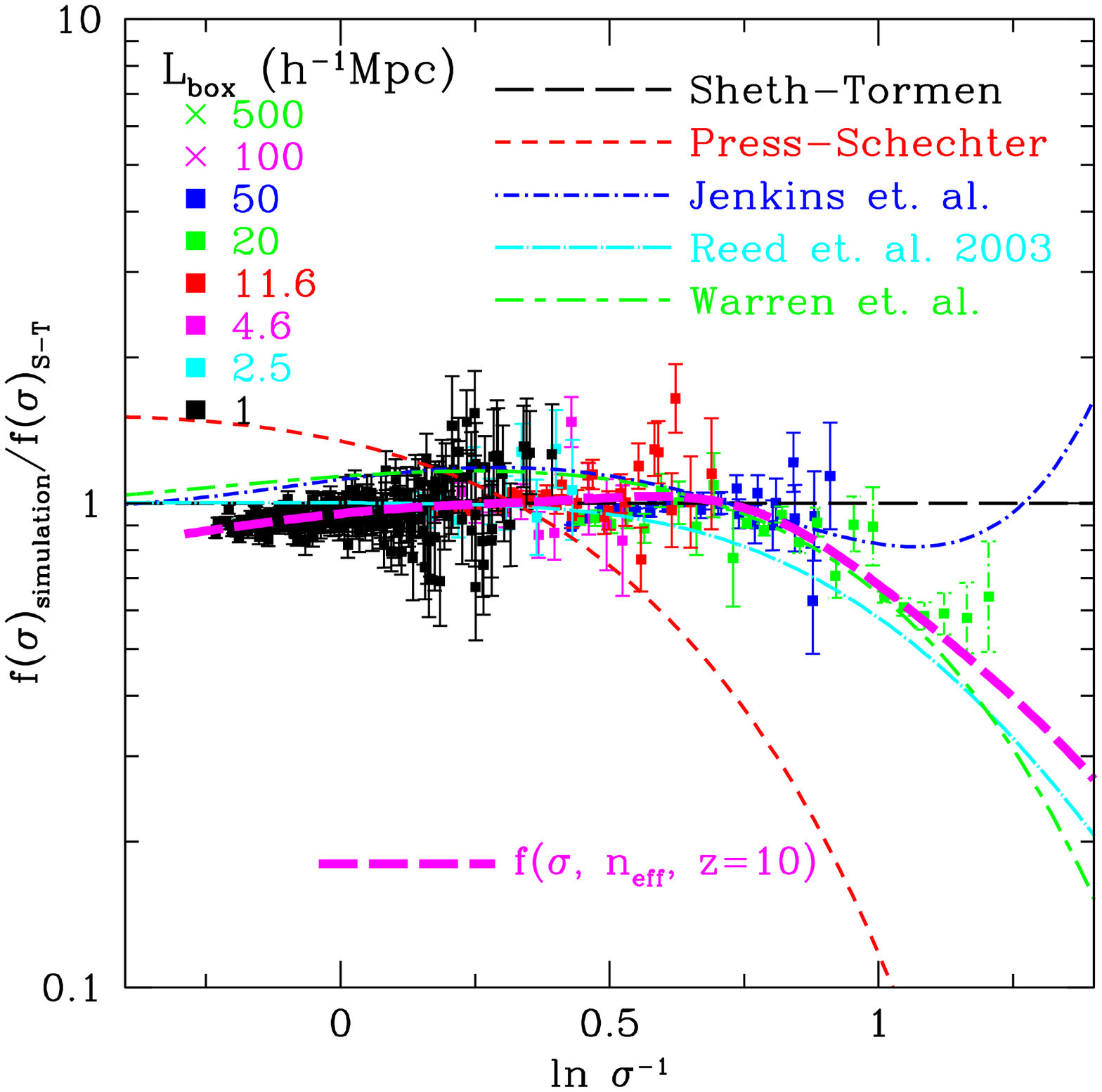, width=0.49\textwidth}

$\begin{array}{c@{\hspace{0.4\textwidth}}c}
\mbox{\bf (a) z$\le$4} & \mbox{\bf (b) z$=$10} 
\end{array}$

   \epsfig{file=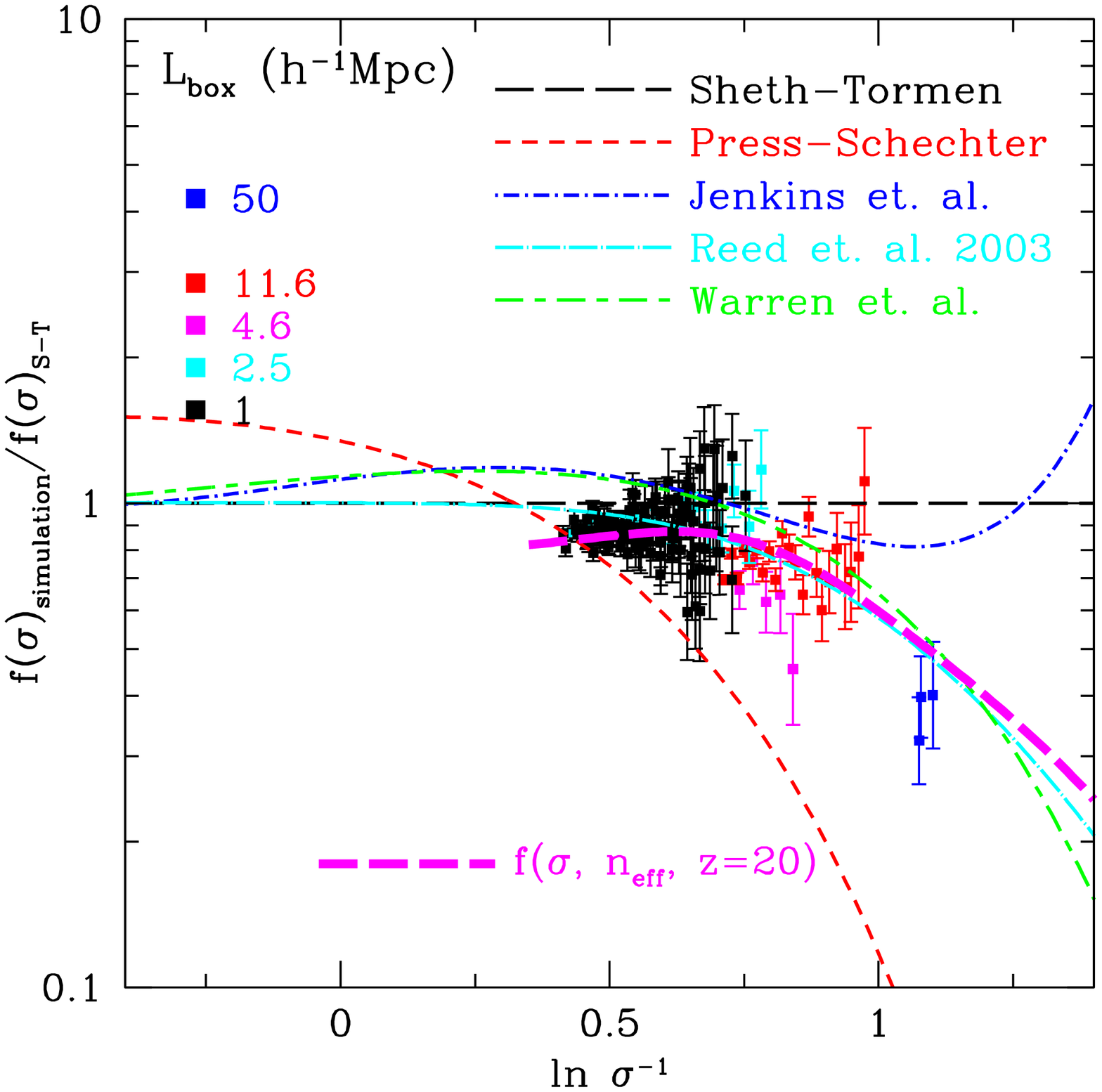, width=0.49\textwidth}   
   \epsfig{file=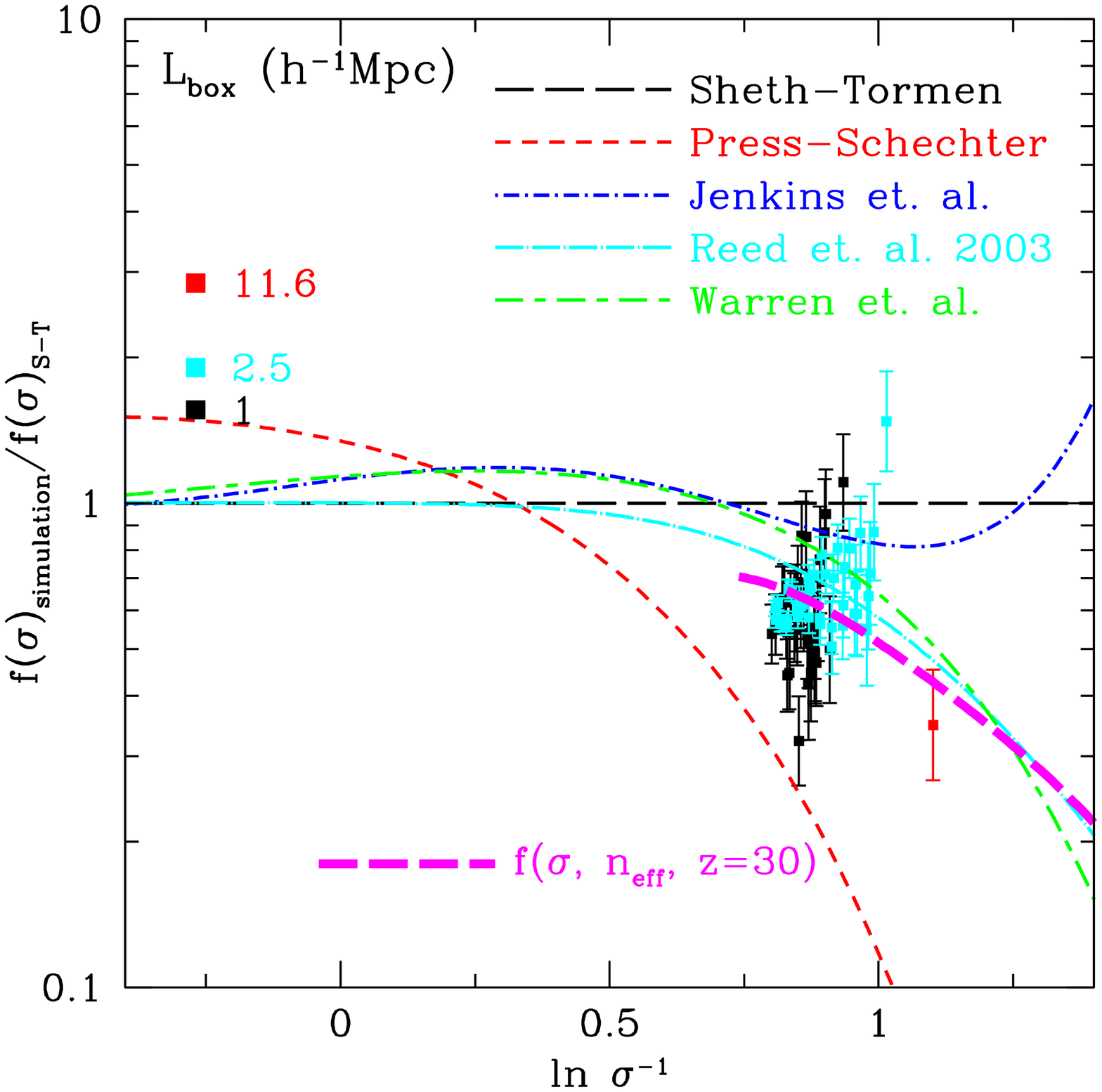, width=0.49\textwidth}

$\begin{array}{c@{\hspace{0.4\textwidth}}c}
\mbox{\bf (c) z$=$20} & \mbox{\bf (d) z$=$30} 
\end{array}$
 
\caption{Differential global (corrected for finite volume; see text)
mass function of {\it friends-of-friends} dark matter haloes for
redshift 10, 20, and 30 compared with analytic fits, including our new
two-parameter fit (Eqn. \ref{neffeqq}), which includes a dependence on
spectral slope, $n_{\rm eff}$, and the resulting redshift dependence. }

\label{nefffits}
\end{center}
\end{figure*}

\begin{figure*}
\epsfig{file=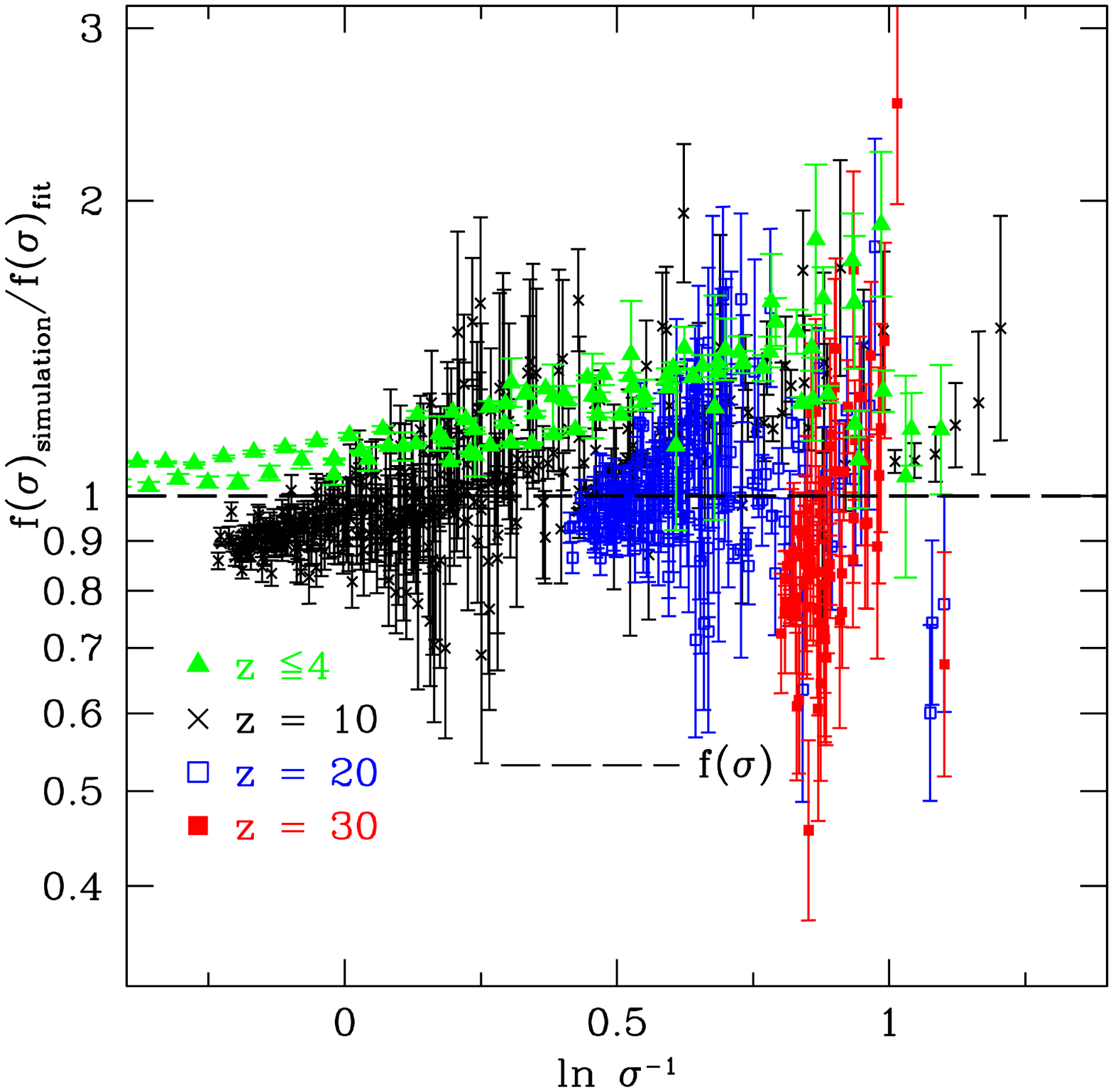, width=0.49\textwidth}
\epsfig{file=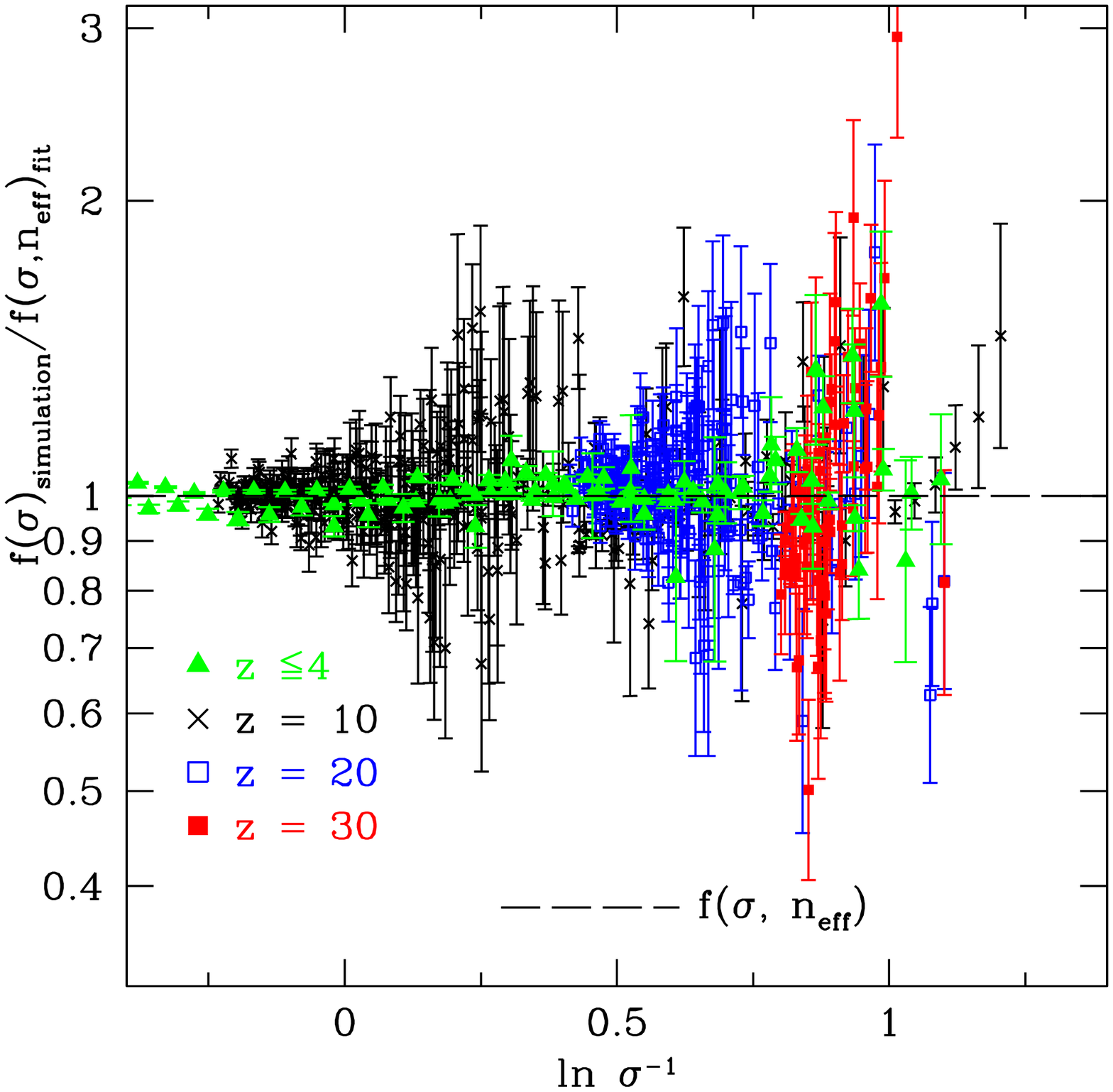, width=0.49\textwidth}

$\begin{array}{c@{\hspace{0.4\textwidth}}c}
\mbox{\bf (a) $f(\sigma)$} & \mbox{\bf (b) $f(\sigma, n_{\rm eff})$} 
\end{array}$

\caption
{Ratio between the differential global (corrected for finite volume;
see text) mass function of {\it friends-of-friends} dark matter haloes
at redshift 0, 1, 4, 10, 20 and 30, and our single parameter fit
(left; Eqn. \ref{stmod1param}); ratio between the same simulation data
and a new fit that includes $n_{\rm eff}$ as a second parameter
(right; Eqn. \ref{neffeqq}).  The improved fit in the right panel shows
that our results are better described by a two-parameter function,
rather than simply a function of $\ln \sigma^{-1}$.  This is evidence
for a weak redshift dependence (via $n_{\rm eff}$) of the mass
function.  }

\label{neffresid}
\end{figure*}

An analytic form for the mass function is an essential ingredient for
a wide array of models of galaxy formation, reionization, and other
phenomena, and is also required for cosmological studies based on
observable objects whose number density depends on the halo mass
function.  In Fig.~\ref{1fit}, we show our data along with several
analytic functional fits; see also Fig.~\ref{nefffits}a-d, where we
plot the mass function split by redshift.  The error bars in
Figs.~\ref{1fit}-\ref{neffresid} are obtained by computing the square
root of the number of haloes in each mass bin  
of a given simulation (see Appendix B for further discussion of 
uncertainty estimates).

The S-T function provides a reasonable fit except for rare haloes
(large $\delta_c/\sigma(m,z)$), where the simulations produce
$\sim$50$\%$ fewer objects.  The P-S function is a poor fit at all
redshifts.  Of the previously published fits, Reed \etal (2003) is the
most consistent with our combined high and low redshift data, fitting
the data with an rms difference of 11$\%$, \ie $\chi^{2}=1$ if we
artificially set the uncertainties to be equal to the Poisson errors
plus 11$\%$ of the measured abundance, added in quadrature.  The
Jenkins \etal function is an excellent fit to our low redshift
simulation data, but it matches the high redshift data less well. This
comparison, however, requires extrapolating the function beyond its
intended range of validity, namely the original fitted range of $-1.2
< \ln \sigma^{-1} < 1.05$.
\footnote{Due to differences in binning the data in the regime where
the mass function is steep, we find the $z=10$ mass function in the
Millennium run (the six rightmost $z=10$ points in Fig. \ref{1fit})
to be $\sim10-20\%$ lower than in Springel \etal (2005), who
found somewhat better agreement with the Jenkins \etal fit.}  The
Warren \etal (2006) curve, which is very similar to the Jenkins \etal
form over its original fitted range, fits our lowest redshift data quite well,
but it is not as good a fit to our high redshift results.

We now consider whether our data support an improved fit compared to
published analytic mass functions. We define the effective slope,
$n_{\rm eff}$, as the spectral slope at the scale of the halo, where
$P(k_h) \propto k_h^{n_{\rm eff}}$ and $k_h=2\pi/r_{0}$, with $r_{0}$
the radius that would contain the mass of the halo at the mean cosmic
density. If we limit the fit to a redshift independent form, with the
assumption of no dependence on $n_{\rm eff}$, our simulation data can
be fit by steepening the high mass slope of the S-T function
(Eq. \ref{stfunc}) with the addition of a new parameter, $c=1.08$, in
the exponential term, and simultaneously including a Gaussian in
$\ln\sigma^{-1}$ centred at $\ln\sigma^{-1}=0.4$, as described by the
following function, which is otherwise identical to the S-T fit:
\begin{eqnarray}\label{stmod1param}
& f(\sigma) = A\sqrt{{2a\over\pi}}
\bigg[1+\big({\sigma^2\over a\delta_c^2}\big)^p + 0.2G_1\bigg]
{\delta_c\over\sigma}\exp\bigg[-{ca\delta_c^2\over2\sigma^2}\bigg]\\
& G_1 = \exp\bigg[-{{[\ln(\sigma^{-1})-0.4]^2} \over {2(0.6)^2}}\bigg] \nonumber.
\end{eqnarray}
For analytical modelling purposes, 
it is useful to recast this equation in terms of a new
variable $\omega^2 = ca\delta_c^2/\sigma^2$
\begin{eqnarray}\label{stmod1param}
& f(\sigma) = A' \omega \sqrt{{2 \over \pi}}
\bigg[1+1.02\omega^{2p} + 0.2G_1\bigg]
\exp\bigg[-{\omega \over 2}\bigg]~~\\
& G_1 = \exp\bigg[-{{[\ln({\omega})-0.788]^2} \over {2(0.6)^2}}\bigg] \nonumber,
\end{eqnarray} 
where $A'=0.310$ and $ca=0.764$.
The resulting function is comparable to the Reed \etal (2003) fit, 
and is generally consistent with the Sheth \& Tormen 2002 
modification to the S-T function with a$=$0.75 instead of 
a$=$0.707 (not plotted).
Note that this modification means that the original normalisation
criteria -- that all mass be contained in haloes,
(Eqn. \ref{integcond}) -- is not satisfied exactly; instead, 98$\%$ of
mass is contained in haloes.  It is remarkable that our data at all
redshifts over a vast range in masses are generally consistent with a
single functional fit that is solely a function of the variance,
independent of redshift.  However, while this redshift independent
function appears reasonable at high redshift, it is relatively poor at
$z=0$.

Careful inspection reveals tentative evidence for a dependence on some
additional free parameter(s).  The mass function at $z \ge 10$ is
suppressed, at levels of $\simgt 10-20\%$, relative to lower
redshifts, indicating a weak dependence of the mass function on
redshift. However, since a given value of $\sigma$ corresponds to
different masses at different redshifts, it is unclear whether the
apparent trend with redshift masks a dependence on mass or on some
other parameter.  Regardless of the cause, inclusion of an additional
parameter in the mass function provides a better fit to our data, as
we now show. We consider the possibility that the mass function may be
affected by $n_{\rm eff}$, the power spectral slope at the scale of
the halo radius.  An improved fit can be made at each redshift with
the introduction of $n_{\rm eff}$ in the analytic function, as given
by the following formula, again a modification to the S-T function:
\begin{eqnarray}\label{neffeqq}
&  f(\sigma, n_{\rm eff}) = A\sqrt{{2a\over\pi}}
\bigg[1+\big({\sigma^2\over a\delta_c^2}\big)^p + 0.6G_1 + 0.4G_2\bigg]
\\
& \times {\delta_c\over\sigma}
\exp\bigg[-{ca\delta_c^2\over2\sigma^2}
-{0.03 \over (n_{\rm eff}+3)^2} \big({\delta_c \over \sigma}
\big)^{0.6} \bigg],\nonumber\\
& G_1 = \exp\bigg[-{{(\ln\sigma^{-1}-0.4)^2} \over {2(0.6)^2}}\bigg],
\nonumber\\
& G_2 = \exp\bigg[-{{(\ln\sigma^{-1}-0.75)^2} \over {2(0.2)^2}}\bigg]
\nonumber
\end{eqnarray}
where $c=1.08$, and $G_1$ and $G_2$ are gaussian functions in
$\ln\sigma^{-1}$. 
This can be rewritten for the purpose of more convenient modelling as
\begin{eqnarray}\label{neffeqqre}
&  f(\sigma, n_{\rm eff}) =
\bigg[1+1.02 \omega^{2p} + 0.6G_1 + 0.4G_2\bigg]
\\
& \times A' \omega \sqrt{2\over\pi}
\exp\bigg[-{\omega \over 2}
-{0.0325 \omega^{p} \over (n_{\rm eff}+3)^2} \bigg],\nonumber\\
& G_1 = \exp\bigg[-{{[\ln({\omega})-0.788]^2} \over {2(0.6)^2}}\bigg] \nonumber,\\
& G_2 = \exp\bigg[-{{[\ln({\omega})-1.138]^2} \over {2(0.2)^2}}\bigg] \nonumber,
\end{eqnarray}
where $A'=0.310$ and $ca=0.764$.  This function fits the data to 4$\%$ rms accuracy,
significantly better than the 15$\%$ rms accuracy of the single
parameter fit of Eqn. \ref{stmod1param}.

The new analytic mass function is presented for redshifts zero through
thirty in Fig. \ref{nefffits}a-d.  The broad ``bump'' over the S-T
function in the $z=0$ mass function centred near $\ln
\sigma^{-1}=0.4$, which is also present in the Jenkins \etal and
Warren \etal fits, is produced by the Gaussian functions in $\ln
\sigma^{-1}$ space.  The $n_{\rm eff}$ term introduces a redshift
dependence that increasingly suppresses the mass function as $n_{\rm
eff}$ approaches -3, and becomes stronger for rarer haloes.  From
Eqn. \ref{varinf}, it is easy to show that for a pure power-law
fluctuation power spectrum, $\sigma^2 ~\alpha ~M^{-(n_{\rm
eff}+3)/3}$, which can be reparameterized as

\begin{equation}
      n_{\rm eff} = 6 {{\rm d}\ln\sigma^{-1}\over{\rm d}\ln
      M\phantom{+}} -3.
\label{defneff}\end{equation}
At fixed redshift, $n_{\rm eff}$ is thus a proxy for halo mass.  At
the smallest scales, the CDM power spectrum asymptotes to $n_{\rm eff}=-3$
for a primordial spectral slope $n_s=1$ We have computed $n_{\rm eff}$
using Eqn. \ref{defneff} throughout this paper.  However, for
convenience, since $n_{\rm eff}$ is nearly linear with $\ln
\sigma^{-1}$ over relatively small ranges in $M$, $n_{\rm eff}$ can be
approximated to better than $10\%$ in ($n_{\rm eff}+3$) by the
following function within the mass and redshift range of haloes in
this paper and for $\sigma_8=0.9$:
\begin{eqnarray}
n_{\rm eff} & \simeq & m_z \ln \sigma^{-1} + r_z \\ m_z & = & 0.55 -
0.32\bigg[1 - \big({1 \over 1+z}\big)\bigg]^5 \nonumber\\ r_z & = &
-1.74 - 0.8\bigg|\log \big({1 \over 1+z }\big) \bigg|^{0.8} \nonumber.
\label{neffeq}
\end{eqnarray}
In Fig.~\ref{neffresid}, we plot the ratio of the simulation data to
the new analytic fits, with (panel~b) and without (panel~a) the
$n_{\rm eff}$ dependence.  The better fit obtained when $n_{\rm eff}$ is
included suggests that the halo mass function is not redshift
independent and thus cannot be described solely by the single
parameter $\sigma(m,z)$. However, panel~a) shows that any dependency
on additional parameters is very weak.
Nevertheless, the precise causes of this apparent dependency 
warrant further study.  
Interestingly, the form of $n_{\rm eff}$ dependence modelled in
peaks theory (\eg Sheth 2001) predicts a smaller difference between
the numbers of high redshift and low redshift haloes than we find in
our simulations. 

\section{sensitivity to cosmological parameters} Our general results
are unaffected by the exact values of the cosmological parameters
because the fit of the mass function in the simulations to an analytic
form is, in principle, independent of the precise relation between
variance and mass (although a dependence on $n_{\rm eff}$ introduces a
weak dependence on cosmological parameters through the relation
between $n_{\rm eff}$ and $\sigma(m,z)$).  Studies involving a wide
range of cosmological parameters (\eg Jenkins \etal 2001; White 2002)
have ruled out a strong dependence of f($\sigma$) on matter or energy
density.  This allows one to estimate the halo abundance for a range
of plausible cosmological parameters using purely the analytic mass
function determined by $\sigma(m,z)$.  In particular, the third year
WMAP results (WMAP-3), which confirm the analysis by Sanchez \etal
(2006) of the first year WMAP and other CMB experiments combined with
the 2dFGRS, imply a fluctuation amplitude significantly smaller than
is commonly assumed, and also suggest a spectral index smaller than 1.
Both of these parameters have a significant impact on the number of
small haloes at high redshift.

\begin{figure}
\epsfig{file=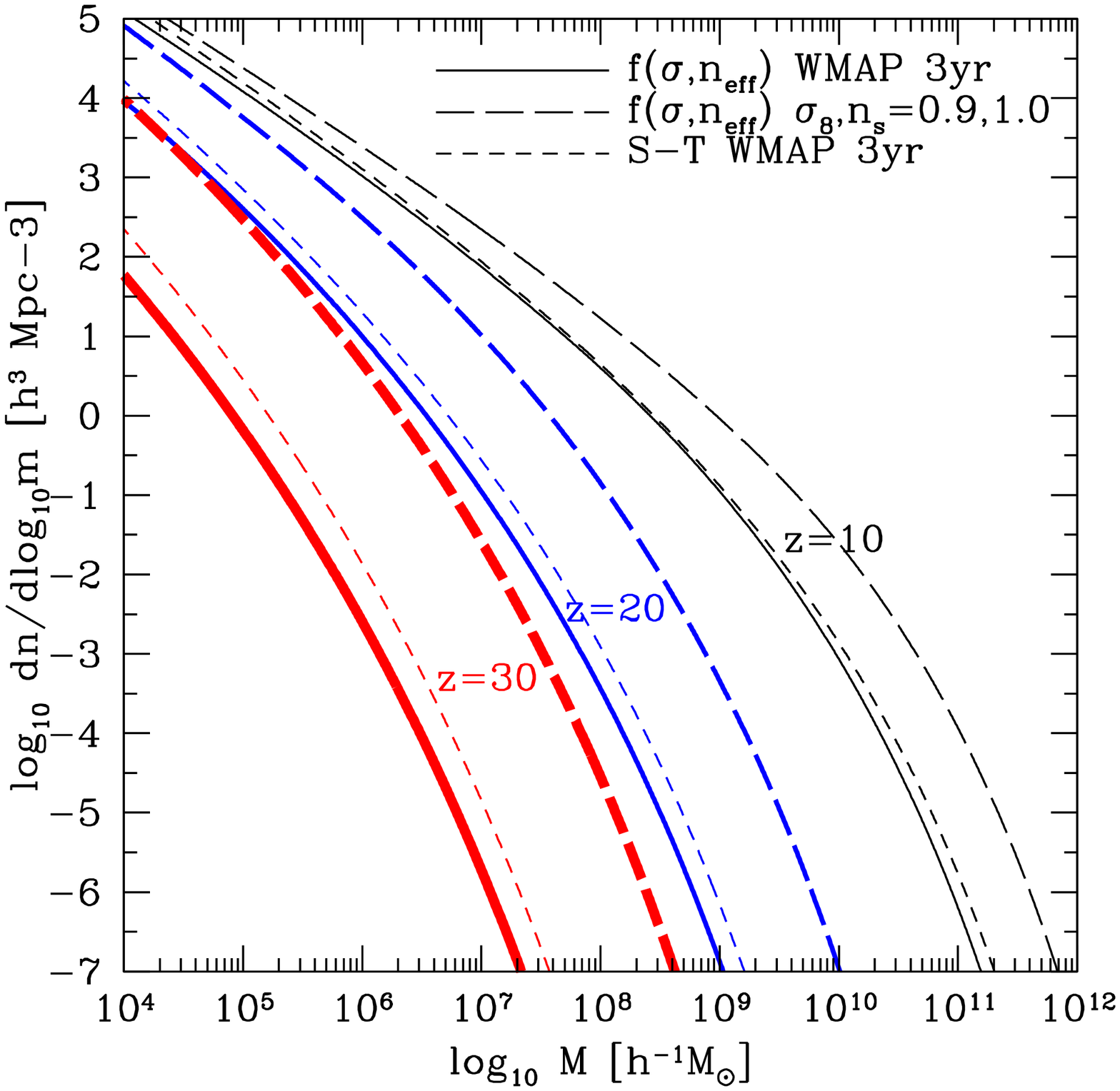, width=\hsize}
\caption{Differential analytic mass function for the WMAP 3-year
cosmological parameters (Spergel \etal 2006), compared with the mass
function for the standard cosmological parameters assumed in the
simulations of this work.  The curves marked ``WMAP 3yr'' use the
preferred WMAP 3-year parameters (\eg $\sigma_8,n_s=0.74,0.951$),
including the effect of the preferred parameters(\eg $\Omega_m$,
$\Omega_{baryon}$, etc) on the transfer function.  The abundance of
massive, high redshift haloes is highly sensitive to cosmological
parameters due to the steepness of the mass function.  }
\label{cosmo}
\end{figure}

\begin{figure}
\epsfig{file=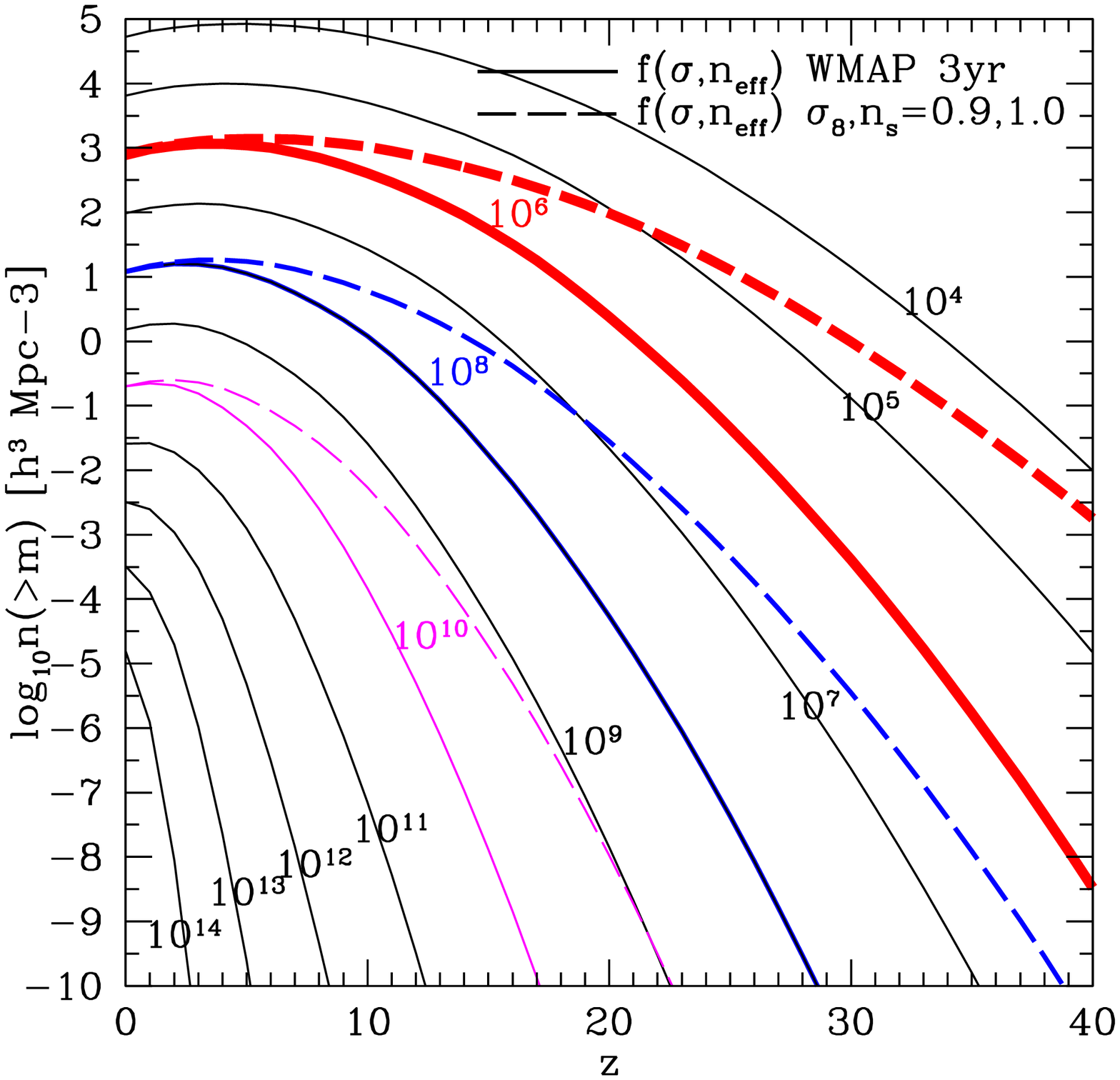, width=\hsize}
\caption{Cumulative abundance versus redshift for the WMAP 3-year
cosmological parameters (Spergel \etal 2006) implied by our new
analytic mass function (Eqn. \ref{neffeqq}). Also shown is the
abundance of haloes for the standard cosmological parameters assumed
in the simulations of this work for 3 masses.  The curves are labelled
by mass in units of $h^{-1}\msun$. The curves marked ``WMAP 3yr''
assume the complete preferred WMAP 3-year parameters
($\sigma_8,n_s=0.74,0.951$), including the effect of the preferred
parameters (\eg $\Omega_m$, $\Omega_{baryon}$, etc) on the transfer
function.  The abundances at high redshift are most sensitive to
cosmological parameters.  }
\label{cosmovsz}
\end{figure}

Figs.~\ref{cosmo} and~\ref{cosmovsz} show that, compared to the
cosmology assumed in the rest of this paper ($\sigma_8=0.9$,
$n_s=1.0$, $\Omega_m=0.25$), the cosmological parameters inferred from
the WMAP-3 data imply a factor of 5 decrease in the number of
candidate galaxy hosts at $z=10$ with mass $\sim10^{8}$ $h^{-1}\msun$,
and more than four orders of magnitude decrease in the number of
potential galaxies at $z=30$ with mass $\sim2\times10^{7}$
$h^{-1}\msun$ .  Smaller ``mini-haloes'' which could host population
III stars are also strongly affected.  The WMAP-3 cosmology implies
less than half the number of ``mini-haloes'' of mass $\sim 10^{6}$ $h^{-1}\msun$ at
$z=10$ and a reduction by more than three orders of magnitude at $z=30$
relative to our standard cosmology.  Note that the comoving abundances
in Fig.~\ref{cosmo} do not match exactly the values in
Fig.~\ref{dnplot} because a slightly different transfer function was
used for some of the simulations, as discussed in \S2.1.

The effect of the WMAP-3 cosmological parameters can also be
interpreted either as reducing the mass of typical haloes at a given
redshift, or as introducing a delay in the formation of structure.
For example, haloes with number density $1 ~h^{3}$Mpc$^{-3}$ at $z=10$
would have a mass approximately four times larger in our standard
cosmology than in the WMAP-3 cosmology; for the same fixed number
density, this becomes a factor of $\sim$10 at $z=20$ and a factor of
$\sim$25 at $z=30$.  For haloes of a fixed mass, the reduced $\sigma_8$
and $n_s$ of the WMAP-3 cosmology delay halo formation.  For example,
haloes of mass of $10^{8} h^{-1}\msun$ are delayed by
$\Delta~z\simeq6$ (z$=$21 to z$=$15) before reaching a comoving
abundance of $10^{-2} ~h^{3}$Mpc$^{-3}$.  Haloes of $10^{6}
h^{-1}\msun$, approximately the mass where H$_{2}$-cooling becomes
strong enough to trigger star formation, do not reach an abundance of
$1 ~h^{3}$Mpc$^{-3}$ until $z=21$ for the WMAP-3 cosmology compared to
$z=30$ for our standard cosmology.  This means that widespread
population III star formation and galaxy formation would occur
significantly later if the WMAP-3 cosmological parameters were
correct.

The uncertainties that remain in the values of cosmological parameters
translate into significant uncertainties in the number of high
redshift haloes that can potentially host luminous objects, especially
those haloes that host the first generations of stars and galaxies.
This adds major uncertainty to predictions of the abundance of
potentially detectable haloes in the pre-reionized universe, such as
those modelled in \eg Reed \etal (2005).  However, the sensitivity of
halo number density to cosmological parameters suggests the exciting
prospect of using the number of small haloes at high redshift as a
cosmological probe if future studies are able to establish their
number density accurately.  
Such measurements of high redshift halo numbers would require not only 
extremely sensitive observations, but also 
further theoretical work to better understand the relation between 
observable properties and halo mass. 

\section{discussion}

Our simulations give the mass function of dark matter haloes out to
redshift 30 and down to masses that include the smallest haloes likely
to form stars, ``mini-haloes'' whose baryons collapse through H$_2$
cooling.  Thus, we now have a precise estimate of the mass function of
the haloes that contain all the stellar material at observable
redshifts, and at virtually all the redshifts that are potentially
observable in the foreseeable future. Our fits were obtained using
simulations of the $\Lambda$CDM cosmology for a specific set of
cosmological parameters, but they are readily scalable to other
values.

These haloes may be detected in a numbers of ways.  Large haloes
(those with T$_{\rm vir} > 10^{4}$K), which have the potential to host
galaxies formed by efficient baryon cooling, might be observed out to
the epoch of reionization with the JW Space Telescope, or perhaps even
with current generation infrared observatories.  Smaller haloes may
not be directly observable.  However, their stellar end-products
might, as gamma-ray bursts (\eg Gou \etal 2004), or as supernovae at
redshifts well above the epoch of reionization (\eg Weinmann \& Lilly
2005).

Knowledge of the halo mass function will be important in the
interpretation of data from the Low Frequency Array (LOFAR), the
Square Kilometer Array (SKA) or other rest-frame 21cm experiments
designed to discover and probe the epoch of reionization. It may even
be possible to use the halo mass function to help break degeneracies
between cosmological parameters and the astrophysical modelling of the
first objects.

Our results for the halo mass function at high redshift imply that
predictions for the abundance of dark matter haloes that may host
galaxies, stars, gamma-ray bursts or other phenomena based on the
commonly used Press-Schechter model grossly underestimate the number
of 3~$\sigma$ or rarer haloes.  This includes large galaxies at
$z\sim10$, all haloes large enough to host galaxies at $z \simgt 15$,
and all haloes capable of hosting stars at $z\simgt20$, assuming that
a halo must be at least $\sim10^{8}$ $h^{-1}\msun$ to host a galaxy
and $10^{6}$$h^{-1}\msun$ to form stars.  The P-S function
underestimates the true mass function by a factor $\sim5$ for the
rarest haloes that we have simulated, which applies to galaxy
candidates at $z=30$, and large ($\sim 10^{11}h^{-1}\msun$) galaxies
at $z=10$.  The abundance of mini-haloes likely to host Population~III
stars are underestimated by the P-S function by a factor of at least
two at $z=30$.  Studies that assume the Sheth \& Tormen function are
more robust, but they still suffer from an overestimation of the
numbers of large haloes at high redshift, particularly of large
galaxies at $z\simgt10$, extending to all potential galaxies at
$z\simgt20$, and to all star-forming haloes at $z\simgt30$, reaching a
factor of up to $\sim$3 for the rarest haloes in our simulations.
However, the S-T function overpredicts the number of mini-haloes only
by less than $\sim20\%$ at $\simlt$20 and by $\sim40\%$ at $z=30$.

\section{summary}

We have determined the mass function of haloes capable of sustaining
star formation from redshift~10 to~30, a period beginning well before
reionization and extending to redshifts below those where reionization
occurred according to the WMAP 3-year estimates.  This extends the mass
function to lower masses and higher redshifts than previous work, and
includes the ``mini-haloes'' that probably hosted population III
stars. Our main results may be summarised as follows:

$\bullet$ We have presented a novel method for correcting for the
effects of cosmic variance and unrepresented large-scale power in
finite simulation volumes.  This allows one to infer more accurately
the true global mass function, ultimately allowing the mass function
to be computed to smaller masses.  We have verified the robustness of
this method by carrying out simulations of a wide range of volumes and
mass resolutions and comparing the inferred mass functions for
overlapping halo mass ranges.  By simulating multiple realisations of
identical volumes, we show that the run-to-run scatter in the mass
function, caused by cosmic variance, is minimised by our method.

$\bullet$ Throughout the period $10<z<30$,  the halo mass function is broadly
consistent with the Sheth \& Tormen (1999) model for haloes $3\sigma$ and
below.  For rarer haloes, the mass function drops increasingly below
the S-T function -- by up to $\sim50\%$ for $\sim5\sigma$ haloes.

$\bullet$ Our data are reasonably well fit by a redshift-independent
function of $\sigma(m,z)$, the rms linear variance in top hat
spheres. We provide a 1-parameter fit to the mass function which is a
modified version of the Sheth \& Tormen (1999) formula. However, an
even better fit can obtained if the mass function is allowed to depend
not only on $\sigma(m,z)$, but also on the slope of the primordial
mass power spectrum, $n_{\rm eff}$.  This improvement implies that
that the fraction of collapsed mass does not depend solely on the rms
linear overdensity, as is assumed in Press-Schechter theory. The P-S
formula, in fact, provides a poor fit to must of our data.

$\bullet$ The halo abundance at $z\simgt10$ is highly sensitive to
$\sigma_8$ and $n_s$, parameters recently adjusted downwards in the
reported WMAP 3-year results (Spergel \etal 2006). The new estimates
imply greatly reduced numbers of high redshift haloes of a given mass.
The sensitivity of high-redshift halo numbers to these parameters
suggests their potential as a useful cosmological probe in future.

Code to generate our best-fit halo mass function may be
downloaded from
http://icc.dur.ac.uk/Research/PublicDownloads/genmf\_readme.html

\section*{acknowledgments}
We thank Raul Angulo for providing halo catalogues for the 1340
$h^{-1}$Mpc simulation.  DR is supported by PPARC.  RGB is a PPARC
Senior Fellow.  TT thanks PPARC for the award of an Advanced
Fellowship. CSF is a Royal Society-Wolfson Research Merit Award
holder. The simulations were performed as part of the simulation
programme of the Virgo consortium.  The new simulations introduced in
this work were performed on the Cosmology Machine supercomputer at the
Institute for Computational Cosmology in Durham, England.  We wish to
thank the referee, Ravi Sheth, for constructive and insightful
suggestions.

{}

\appendix
\section{numerical tests}
\subsection{Run parameter convergence tests}

\begin{figure*}
  \begin{center}
 
    \epsfig{file=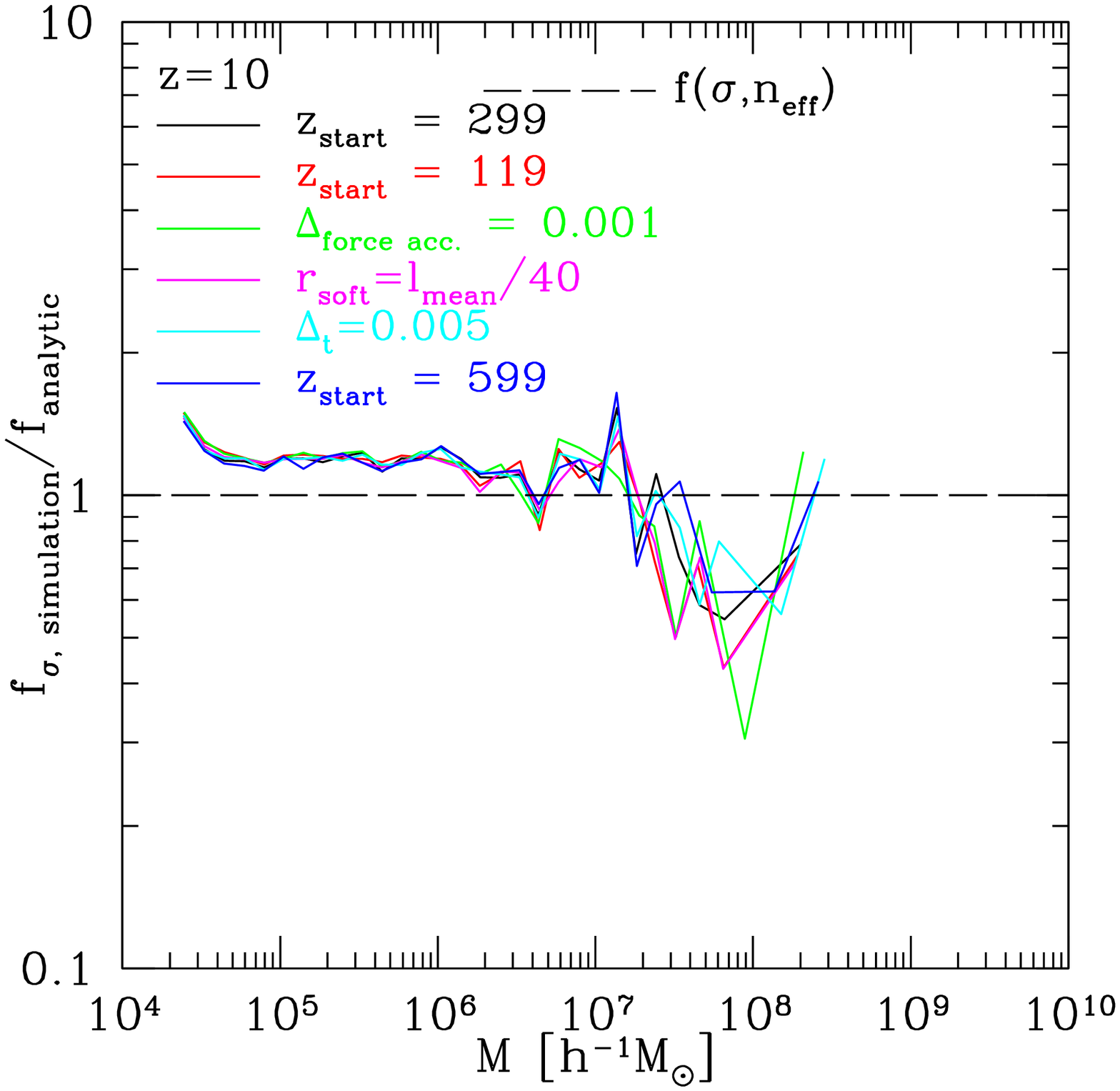, width=0.33\textwidth}
     \epsfig{file=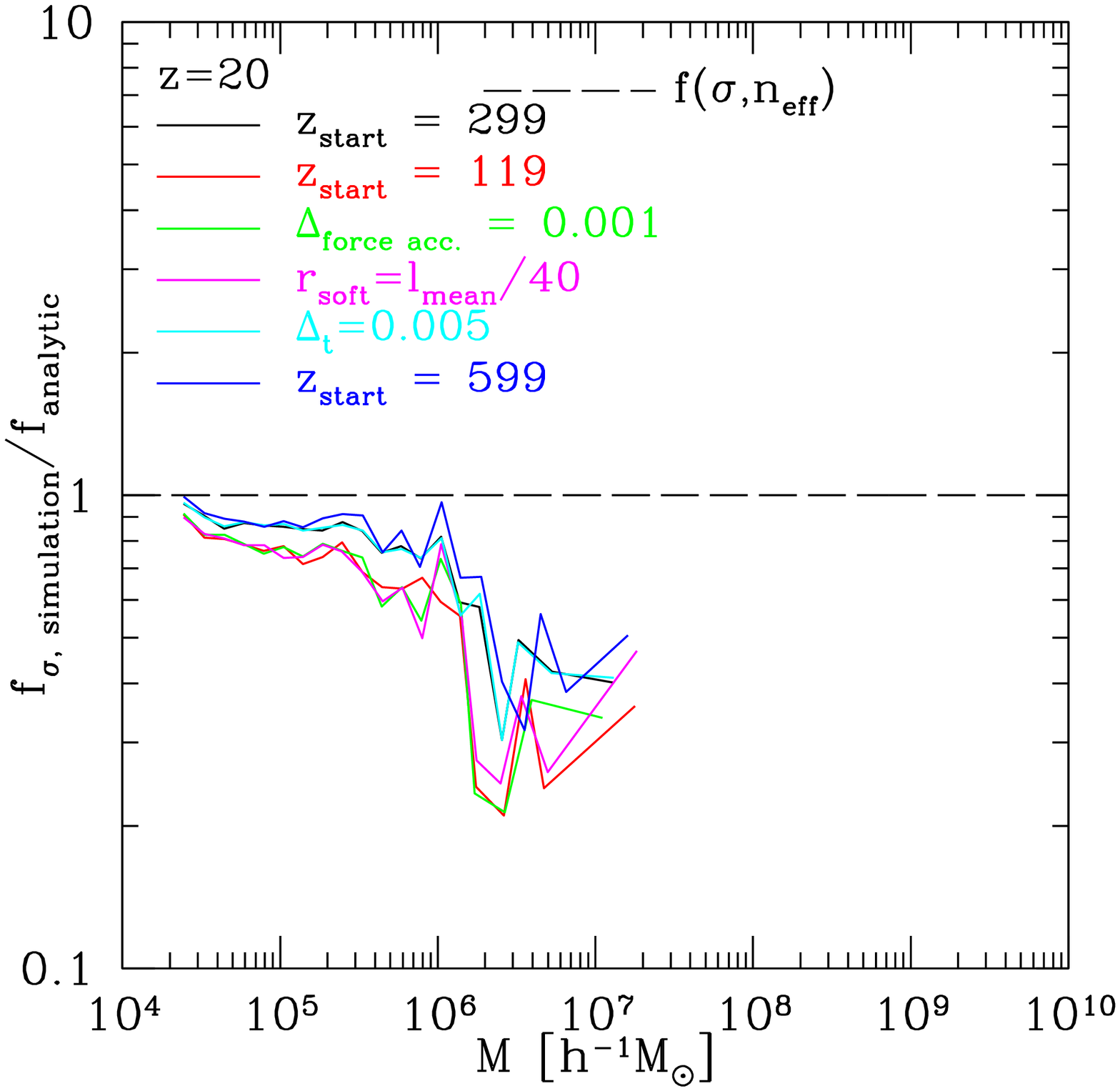, width=0.33\textwidth}
    \epsfig{file=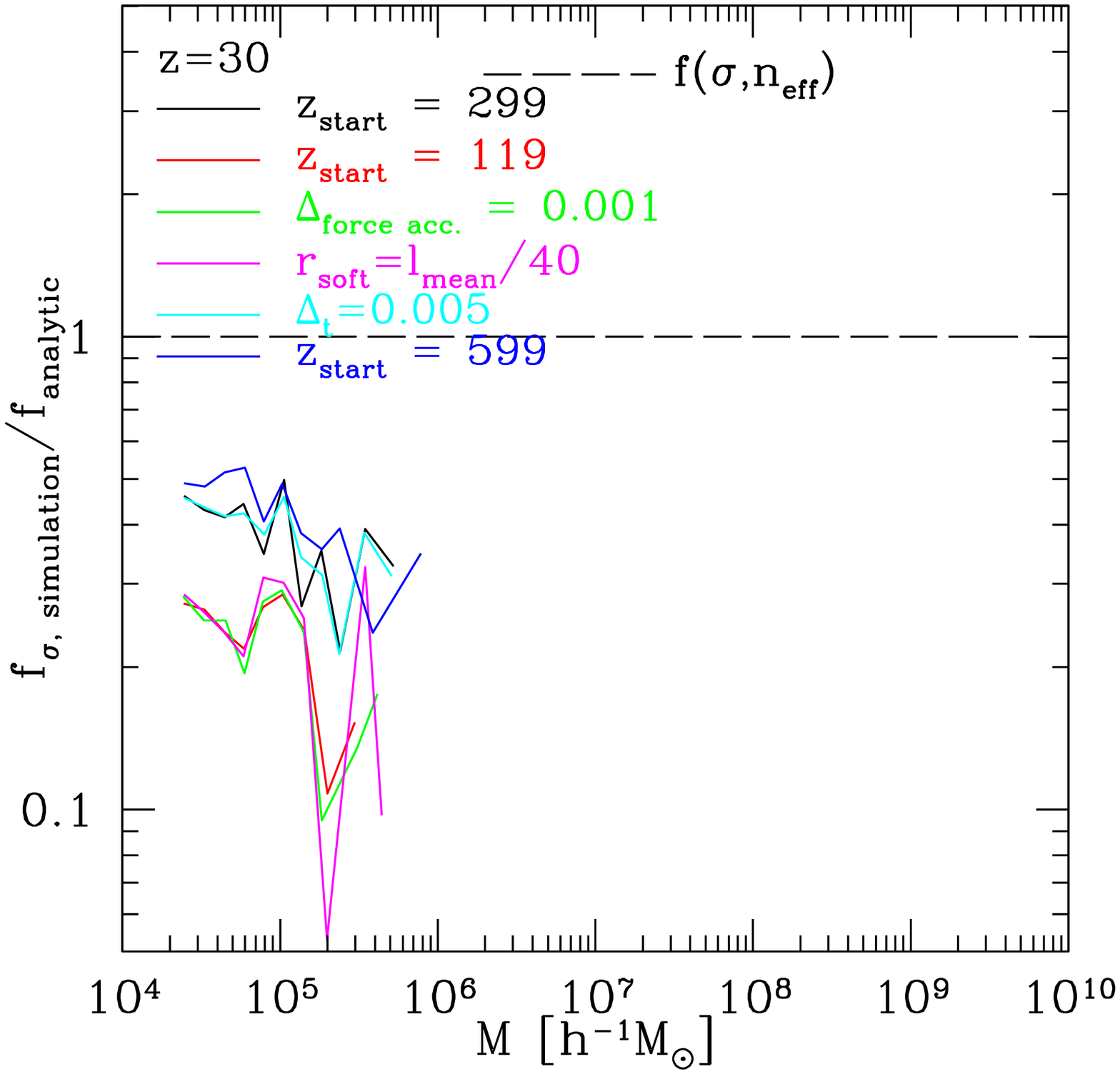, width=0.33\textwidth}

$\begin{array}{c@{\hspace{0.25\textwidth}}c@{\hspace{0.25\textwidth}}c}
\mbox{\bf (a) z$=$10} & \mbox{\bf (b) z$=$20} & \mbox{\bf (c) z$=$30}
\end{array}$
 
\caption{Differential raw (unadjusted for finite volumes) simulated
mass function of {\it friends-of-friends} dark matter haloes for
redshifts 10, 20, and 30 is shown for several run parameters.  Unless
otherwise noted, each run used the parameters implemented throughout
this paper (${\rm z_{start}=119}$, ${\rm \Delta_{force~acc}=0.005}$,
${\rm r_{soft}=l_{mean}/20}$, $\Delta_t=0.02$) , except that the green
(${\rm \Delta_{force~acc}}$), and magenta (r$_{\rm soft}$) curves used
${\rm z_{start}=119}$.  Here haloes are plotted down to 20 particles
per halo versus the 100 particle minimum imposed throughout the paper;
particle mass is $1.1\times10^{3} h^{-1} \msun$.  }
\label{runtests}
\end{center}
\end{figure*}

In Fig. \ref{runtests}, we have tested some of the primary runtime
parameters of L-Gadget2 using a $10^{3}$$h^{-1}\msun$ particle
resolution 1 $h^{-1}$Mpc volume.  These tests confirm that our choices
of starting redshift (${\rm z_{start}}$), fractional force accuracy
(${\rm \Delta_{force~acc}=0.005}$), softening length (${\rm
r_{soft}}$), and maximum allowed timestep
($\Delta_t=\Delta\ln(1+z)^{-1}$) are sufficient.  Of the parameters
that we have tested, ${\rm z_{start}}$ has the most effect on our
results.  At z$=$10, ${\rm z_{start}}=119$ is indistinguishable from
earlier starting redshifts.  However, at z$=$20, the ${\rm
z_{start}}=119$ mass function is suppressed by $\sim10-20\%$ relative
to the ${\rm z_{start}}=299$ runs, and by more than $50\%$ at z$=$30.
At z$=$30, the ${\rm z_{start}}=299$ run is suppressed relative to the
${\rm z_{start}}=599$ run by $\sim10-20\%$, which could indicate a
small bias in our z$=$30 mass function, but not large enough to affect
significantly our conclusions.  It thus appears that a simulation must
be evolved a factor of $\sim$10 in expansion factor in order to solve
accurately (within 10-20$\%$) the mass function for the mass
resolutions and outputs we have considered, if one uses the Zel'dovich
(1970) approximation to set up initial conditions, as we have.
Note that the convergence of our runs with increased 
starting redshift implies that
the Zel'dovich approximation is valid.  See discussion in following section.

\subsection{Resolving haloes}
Studies of the mass function at low redshift have found that the mass
function is adequately sampled for haloes containing as few as 20
particles (\eg Jenkins \etal 2001).  Because we are exploring new
regimes in mass and redshift, it is necessary to confirm that we model
haloes with sufficient particle numbers to resolve the mass function.
The FOF mass function for haloes of very few particles is typically
enhanced artificially as spurious groupings are increasingly common
for small particle numbers.

\begin{figure}
\epsfig{file=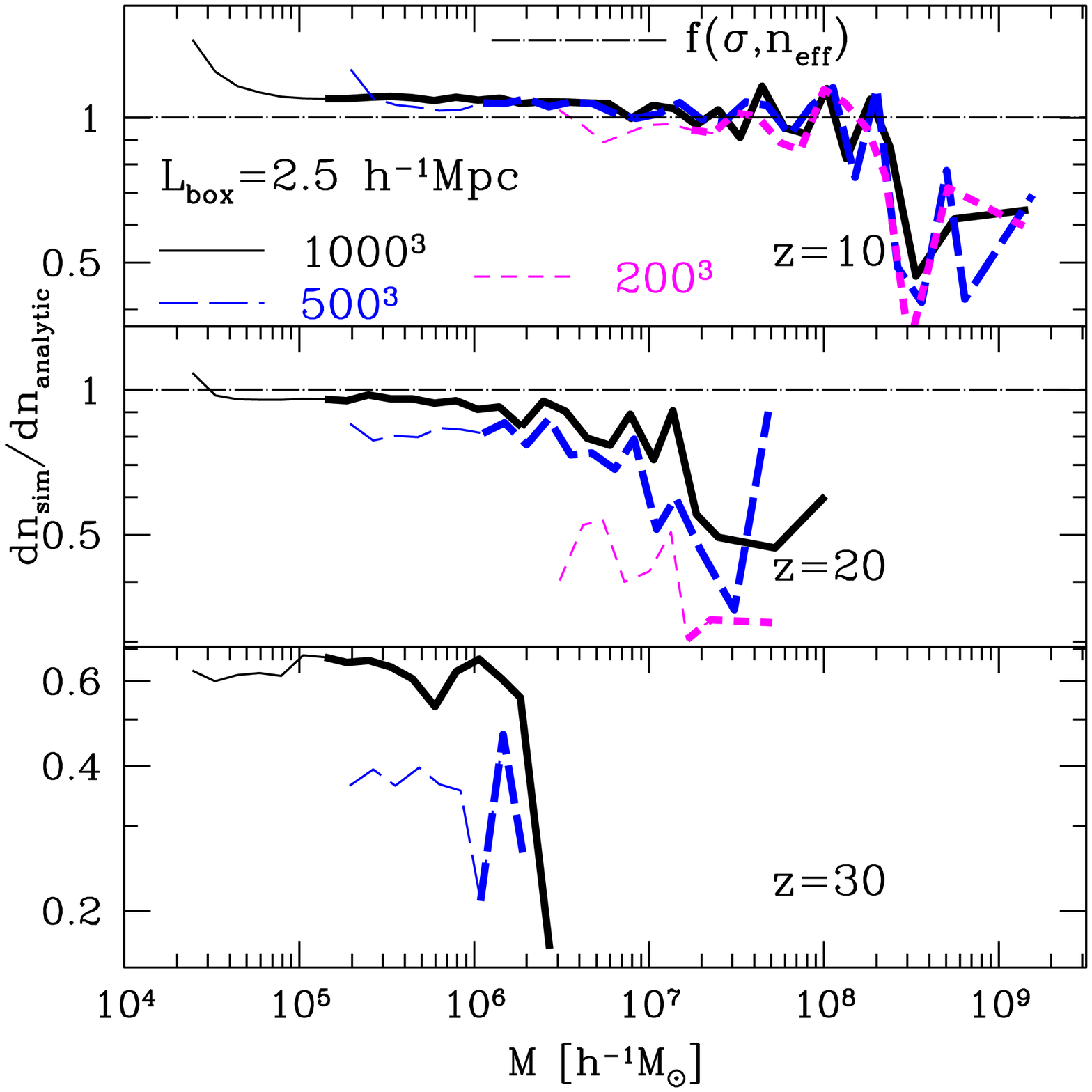, width=\hsize}
\caption{Residuals of the differential raw (unadjusted for finite
volumes) mass function for a 2.5 $h^{-1}$Mpc box with identical
initial density fluctuations modelled at 3 different mass resolutions
(200$^{3}$, 500$^{3}$, and 1000$^{3}$ particles).  Thick line segments
denote haloes of at least 100 particles, the minimum particle number
that we implement throughout the paper.  Thin line segments show the
mass function down to 20 particles per halo.}
\label{halores}
\end{figure}

Our resolution tests consists of an identical volume, simulated at
multiple mass resolutions.  The mass function of a 2.5 $h^{-1}$Mpc box
at resolutions of 200$^{3}$, 500$^{3}$, and 1000$^{3}$ is shown in
Figure \ref{halores}.  At each resolution, the mass function has an
upturn below approximately 30-40 particles.  Additionally, the mass
function is suppressed over the range of $\sim$30 to $\sim$100
particles, at a level that appears to depend on redshift.  At scales
larger than approximately 100 particles, the mass function at multiple
resolutions agrees within the uncertainties.  For this reason, we
limit our analysis to haloes of at least 100 particles.  This
resolution limit is significantly higher than found necessary at low
redshift in many previous works.  Even with this conservative particle
resolution limit, some small bias in the mass function cannot be ruled
out fully for redshifts 20 or higher.  The increased sensitivity of
the mass function to particle resolution as redshift increases is
likely enhanced by the increased steepness of the mass function in
this regime for haloes formed from rare fluctuations.  The suppression
of the mass function for haloes of fewer than 100 particles may be due
to transient effects that become unimportant by lower redshifts.
These effects could include errors introduced as a result of
inaccuracies of the Zel'dovich approximation (Zel'dovich 1970) where
initial particle positions and velocities are computed based on the
initial density field, which is assumed to be entirely linear.
If so, then these transients could be reduced by 
using second order Lagrangian perturbation 
theory (2LPT; Scoccimarro 1998, Crocce, Pueblas \& Scoccimarro 2006) 
to set up initial
conditions.  Use of 2LPT may also reduce the required starting redshift.
Further investigation is required to determine more fully the sources
of increased sensitivity to mass resolution on the mass function.

\subsection{Halo finders: friends-of-friends (FOF) versus spherical overdensity (SO)}
Choice of halo finder can have a major impact on halo masses and on
the mass function.  It is thus worth considering how our results would
change had we used a different halo finder.  In this case, the FOF
mass function is compared with the mass function produced by the {\it
spherical overdensity} SO algorithm (Lacey \& Cole 1994), which
identifies spheres of a specified overdensity.  FOF is computationally
efficient and will select objects of any shape provided that they meet
a local particle density.  However, FOF may also spuriously link
together neighbouring haloes, which is a potential issue for low mass,
high redshift haloes which form at scales where mass fluctuation
spectrum is steep, and result in highly ellipsoidal halo shapes (Gao
\etal 2005).  Because SO assumes haloes are spheres, it is not ideal
for highly ellipsoidal haloes.  However, SO has an advantage over FOF
in that it is less likely spuriously to link together neighbouring
haloes or to misclassify highly ellipsoidal but unvirialized
structures as haloes.

\begin{figure}
\epsfig{file=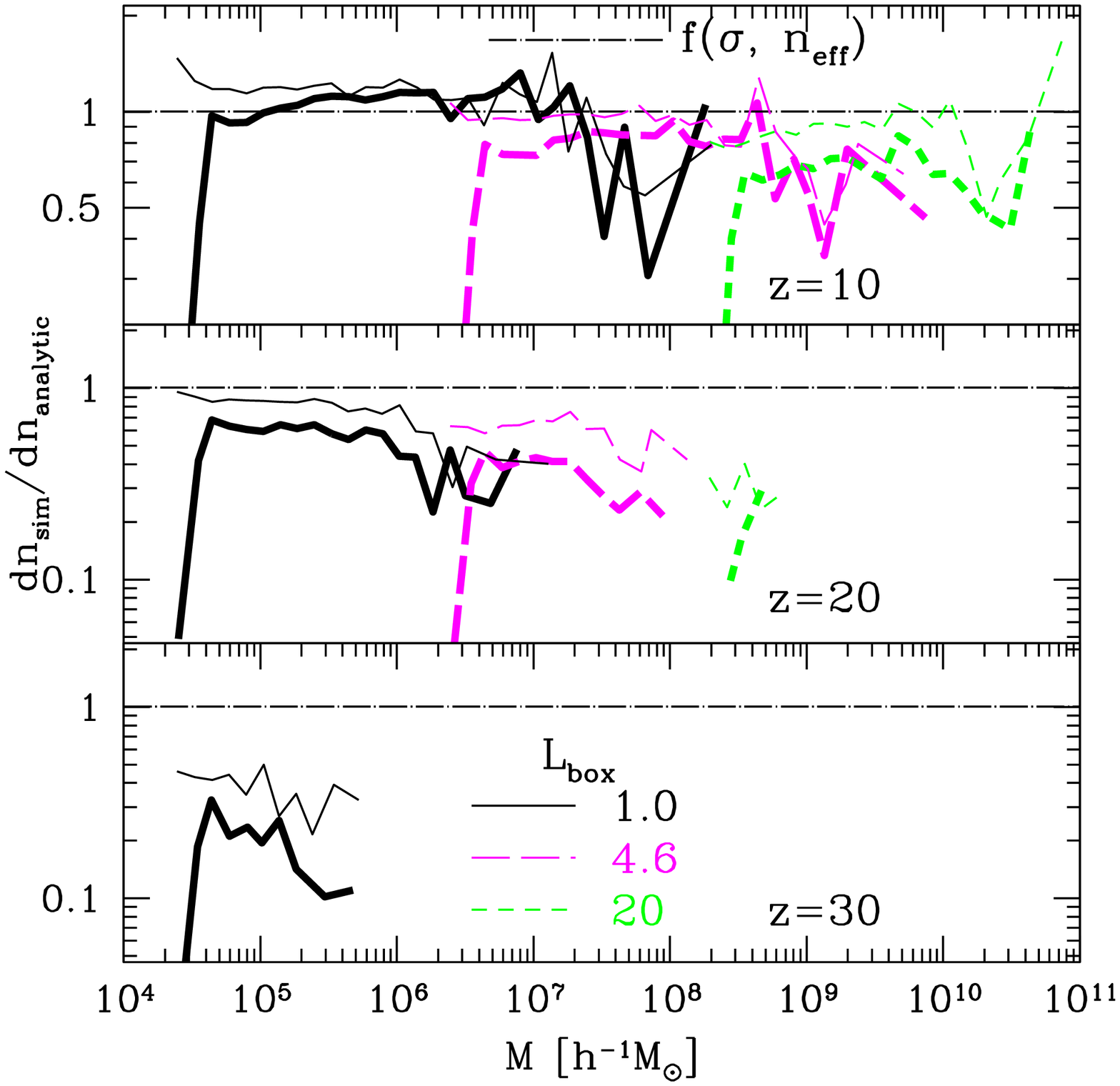, width=\hsize}
\caption{{\it friends-of-friends} (FOF, thin lines) and {\it spherical
overdensity} (SO, thick lines) halo differential raw (unadjusted for
finite volumes) mass functions for particle masses of $\sim$ 10$^3$,
10$^{5}$, and 10$^{7}$ $h^{-1}\msun$ (box size of 1.0, 4.6, and 20
$h^{-1}$Mpc with 400$^{3}$ particles). 
SO density is 178 times the mean density. Curves are
plotted down to 20 particles per halo. }
\label{sofof}
\end{figure}

We have tested the SO algorithm assuming the spherical tophat model,
in which the $\Lambda$CDM virial overdensity, {\it $\Delta_{\rm
vir}$}, in units of the mean density is 178 at high redshifts, when
$\Omega_m\simeq 1$.  We have computed the SO mass function for three
simulations of particle mass resolution $\sim$ 10$^3$, 10$^{5}$, and
10$^{7}$ $h^{-1}\msun$ at redshifts 10, 20, and 30.  In Figure
\ref{sofof}, we show the SO and FOF mass functions for these outputs.
In general, the SO mass function is lower than the FOF mass function,
though the two mass functions are consistent for much of the redshift
10 mass range.  The difference between the two mass functions
increases with mass and redshift, ranging from $\simlt$10$\%$ at
redshift 10 and is generally less than a factor of 2.  Some caution
should be taken when considering the SO mass function because of its
particle number dependence in this implementation of the SO halo
finder.  Initial candidate centres for SO haloes were identified by
finding density peaks, where local density was computed for each
particle, smoothed by its 32 nearest neighbours.  Spheres were then
grown outward until the desired overdensity was reached.  This means
that SO haloes with masses approaching and below the 32 particle
smoothing mass will be suppressed.

While there are some differences in the mass function for the two
means of identifying haloes, these differences are generally smaller
than the differences introduced by adopting the WMAP 3 year
cosmological parameters versus the larger $\sigma_8=0.9$ and $n_s=1.0$
used in the simulations of this paper.

\section{uncertainties and run to run scatter}

\begin{figure}
\epsfig{file=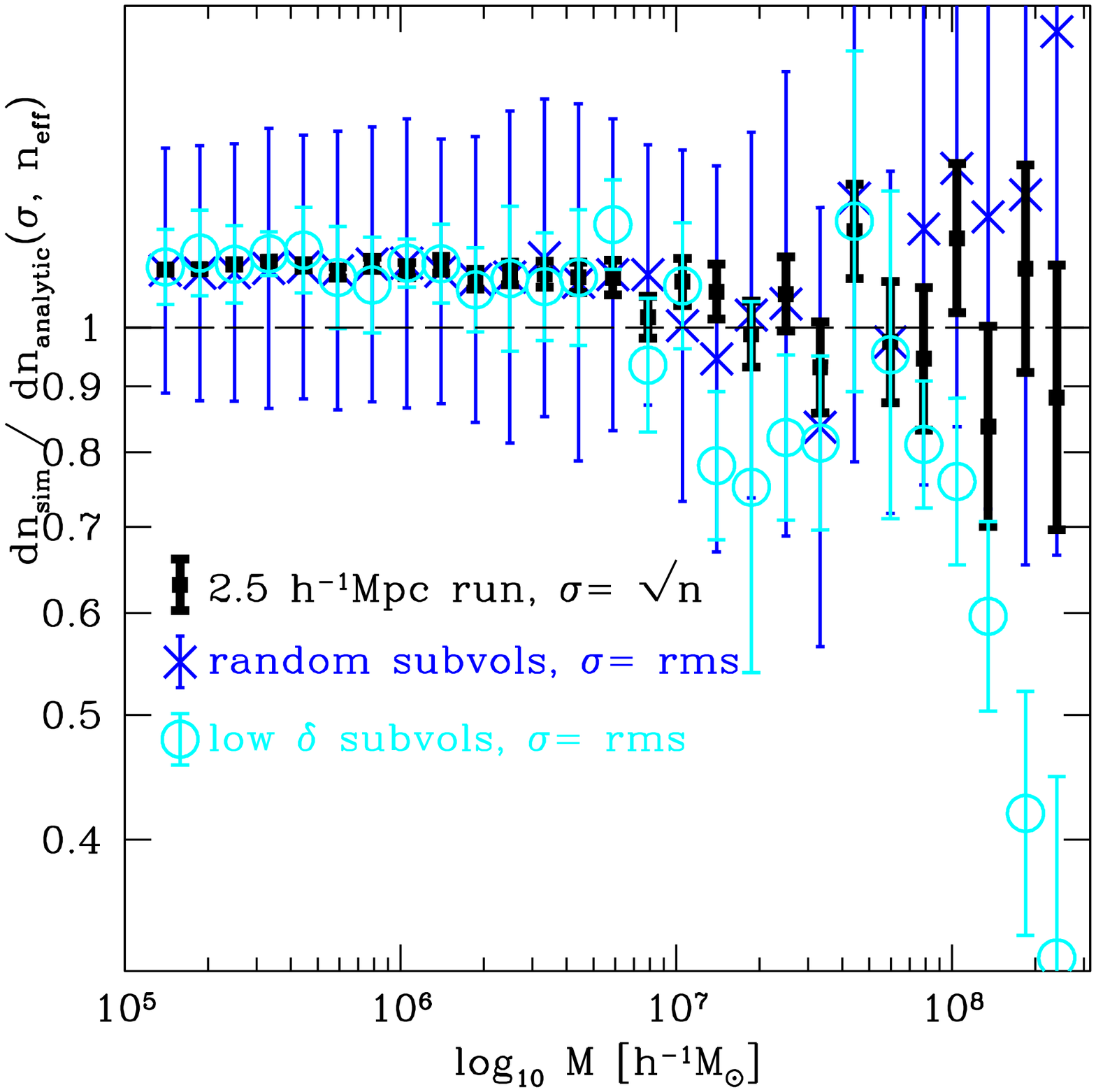, width=\hsize}
\caption{ Comparison of uncertainty estimators using the differential
raw (unadjusted for finite volumes) mass functions for particle masses
of z$=$10.  Filled squares (black) show the 2.5 $h^{-1}$Mpc run with
poisson errors (where uncertainty is assumed to be equal to the square
root of the number of haloes in each mass bin, as throughout the
paper).  Large X's (blue) denote the mean and rms variance of the mass
function of 11 random 1 $h^{-1}$Mpc subvolume cubes selected from the
same volume.  Open circles (cyan) are the mean and rms of 3 low
overdensity 1 $h^{-1}$Mpc subvolmes. }
\label{errssubs}
\end{figure}

\begin{figure}
\epsfig{file=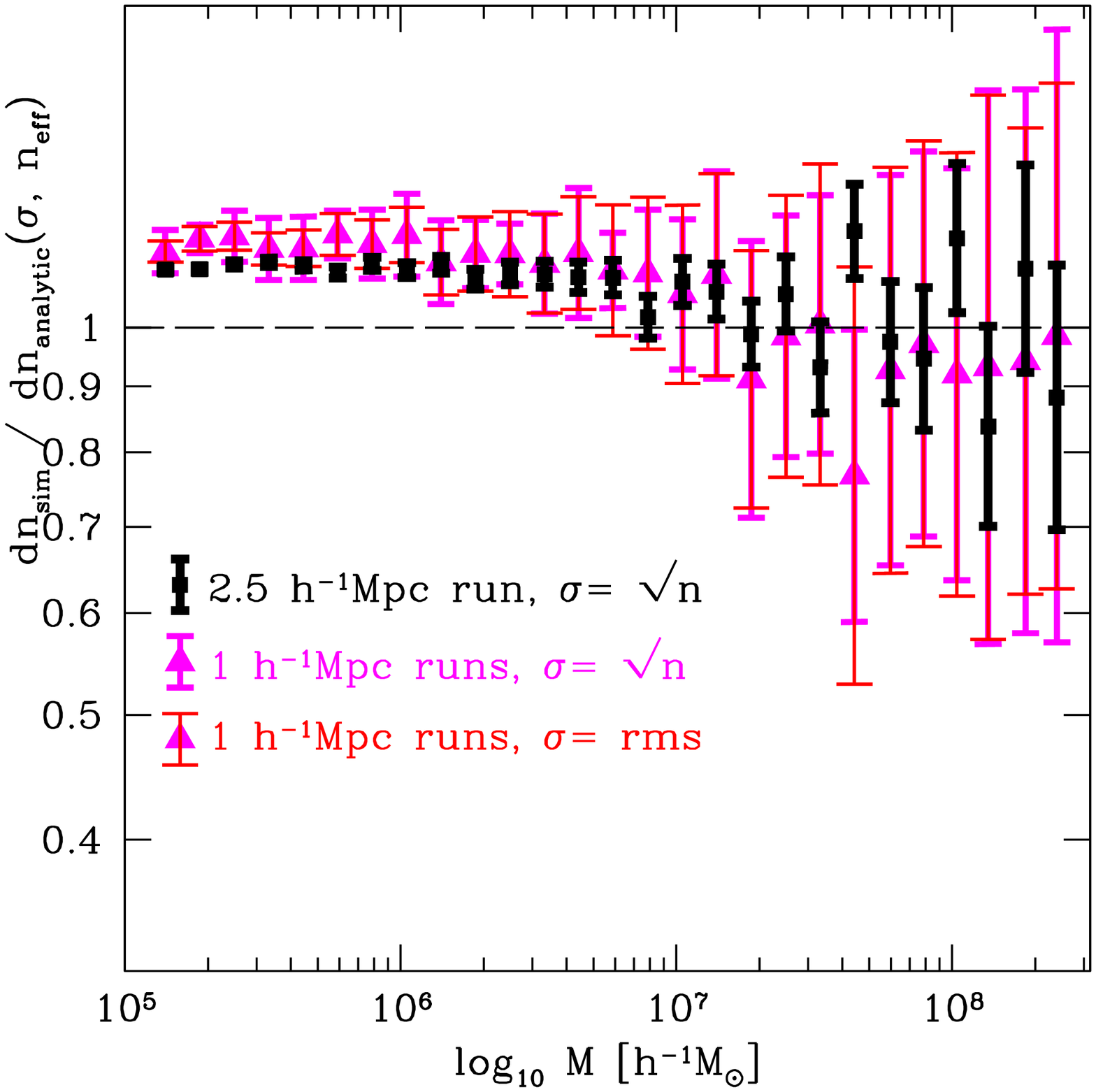, width=\hsize}
\caption{ Comparison of uncertainty estimators using the differential
raw (unadjusted for finite volumes) mass functions for particle masses
of z$=$10.  Filled squares (black) show the 2.5 $h^{-1}$Mpc run with
$\sqrt{n}$ poisson errors.  Triangles (magenta) denote the mean mass
function of the eleven 1 $h^{-1}$Mpc simulations with poisson
uncertainty (magenta error bars) and run to run rms variance (red error
bars). }
\label{errsboxes}
\end{figure}

In this section, we discuss uncertainties and run to run variance
using the 2.5 $h^{-1}$Mpc and 1 $h^{-1}$Mpc boxes at z$=$10 as
examples.  Fig. \ref{errssubs}-\ref{errsboxes} shows that the poisson
($\sqrt{n}$) estimate of uncertainty is much smaller than the
``bootstrap'' uncertainty computed by taking the rms variance between
random subsamples.  The scatter between 3 non-overlapping subsamples
with low overdensity, ($|\delta| < 10^{-3}$), is much smaller, and is
in fact comparable to the scatter between individual simulations of 1
$h^{-1}$Mpc cubes.  Although there are too few low overdensity
subvolumes for a truly representative sample, their similarity
suggests that most of the variance between the random subvolumes is
due to non-zero mean density.

Between the 11 small (1 $h^{-1}$Mpc) simulations, the run to run
variance is comparable to their individual average poisson
uncertainty.  This suggests that poisson uncertainty provides a
reasonable estimate of the true uncertainty provided that other finite
volume effects, such as missing large scale power, are taken into
account (see earlier discussion).  The mean mass function for low mass
haloes is consistent among these subvolumes and small volume
simulations.  However, for the low overdensity subvolumes, the mean
mass function is deficient in massive haloes.  

\label{lastpage}

\end{document}